\documentclass[aps,nofootinbib,amsmath,prc,showpacs,groupedaddress,12pt,superscriptaddress,notitlepage]{revtex4-1}

\usepackage[paperwidth=210mm,paperheight=297mm,centering,hmargin=2.4cm,tmargin=2.8cm,bmargin=3.5cm]{geometry}

\usepackage{amsfonts}
\usepackage{amsmath}
\usepackage{epsfig}
\usepackage{latexsym}
\usepackage{paralist}
\usepackage{fancyhdr}
\usepackage{graphicx}

\numberwithin{equation}{section}

\usepackage[vcentermath]{youngtab}
\usepackage{young}
\usepackage{ytableau}
\usepackage{etex}
\usepackage{braket}
\usepackage{float}
\usepackage{slashed}
\usepackage{xcolor}
\usepackage{subcaption}
\usepackage{soul}

\usepackage{dcolumn}

\makeatletter
\@addtoreset{equation}{section}

\makeatother

\def\bea{\begin{eqnarray}} 
\def\eea{\end{eqnarray}}
\def\be{\begin{equation}} 
\def\ee{\end{equation}} 
\def\ba{\begin{array}}
\def\ea{\end{array}} 
\def\nn{\nonumber}

\def\be{\begin{equation}}
\def\ee{\end{equation}}
\def\bea{\begin{eqnarray}}
\def\eea{\end{eqnarray}}

\def\Tr{\mbox{Tr}}

\def\nn{\nonumber}

\renewcommand{\thefootnote}{\fnsymbol{footnote}}

\let\oldtitle\title
\renewcommand{\title}[1]{\oldtitle{\color{blue}{#1}}}

\let\oldeqref\eqref
\let\oldcite\cite
\renewcommand{\eqref}[1]{{\color{blue}\oldeqref{#1}}}
\renewcommand{\cite}[1]{{\color{blue}\oldcite{#1}}}

\makeatletter
\let\reftagform@=\tagform@
\def\tagform@#1{\maketag@@@{\ignorespaces\textcolor{blue}{(\ignorespaces #1 \unskip\@@italiccorr \ignorespaces)\ignorespaces}}}
\makeatother

\makeatletter
\renewcommand{\p@subsection}{}
\renewcommand{\p@subsubsection}{}
\makeatother

\usepackage{mathpazo}
\usepackage{amsmath}

\begin{document}


\title{Functional perturbative RG and CFT data in the $\epsilon$-expansion}

\author{A.\ Codello}
\email{codello@cp3-origins.net}
\affiliation{CP$^3$-Origins, 
University of Southern Denmark,
Campusvej 55, 5230 Odense M, Denmark}
\affiliation{INFN - Sezione di Bologna, via Irnerio 46, 40126 Bologna, Italy}

\author{M.\ Safari}
\email{safari@bo.infn.it}
\affiliation{INFN - Sezione di Bologna, via Irnerio 46, 40126 Bologna, Italy}
\affiliation{
Dipartimento di Fisica e Astronomia,
via Irnerio 46, 40126 Bologna, Italy}

\author{G.\ P.\ Vacca}
\email{vacca@bo.infn.it}
\affiliation{INFN - Sezione di Bologna, via Irnerio 46, 40126 Bologna, Italy}

\author{O.\ Zanusso}
\email{omar.zanusso@uni-jena.de}
\affiliation{
Theoretisch-Physikalisches Institut, Friedrich-Schiller-Universit\"{a}t Jena,
Max-Wien-Platz 1, 07743 Jena, Germany}
\affiliation{INFN - Sezione di Bologna, via Irnerio 46, 40126 Bologna, Italy}

\begin{abstract}
\vspace{3mm}
We show how the use of {\it standard} perturbative RG in dimensional regularization
allows for a renormalization group based computation
of both the spectrum and a family of 
coefficients of the operator product expansion (OPE) for a given universality class.
The task is greatly simplified by a straightforward generalization of perturbation theory
to a {\it functional} perturbative RG approach.
We illustrate our procedure in the $\epsilon$-expansion by obtaining the next-to-leading corrections 
for the spectrum and the leading corrections for the OPE coefficients of {\tt Ising} and {\tt Lee-Yang}
universality classes and then give several results for the whole family of renormalizable multicritical models $ \phi^{2n}$.
Whenever comparison is possible our RG results explicitly match the ones recently derived in CFT frameworks.
\end{abstract}

\pacs{}
\maketitle

\renewcommand{\thefootnote}{\arabic{footnote}}
\setcounter{footnote}{0}


\section{Introduction}\label{intro}

The standard perturbative renormalization group (RG) and the $\epsilon$-expansion have been,
since the pioneering work of Wilson and Kogut \cite{Wilson:1973jj}, the main analytical tools for the analysis of critical phenomena
and, more generally, for the study of universality classes with methods of quantum field theory (QFT).
Under the pragmatic assumption that scale invariance implies conformal invariance at criticality,
which is confirmed by almost all interesting examples,
one could also argue that conformal field theory (CFT) methods
serve as an additional theoretical tool to describe critical models.

The RG flow of deformations of a scale invariant critical theory in a given operator basis is generally encoded in a set of beta functions of the corresponding couplings.
As demonstrated by Cardy~\cite{Cardy:1996xt},
the beta functions can be extracted adopting a microscopic short distance cutoff as a regulator and in particular, expanding in the scaling operators,
the linear part of the beta functions is controlled by the scaling dimensions of the associated operators,
while the quadratic part is fixed by the OPE coefficients of the operators involved in the expansion.
Whenever the underlying critical model is a CFT, this approach is the foundation of conformal perturbation theory
and its development strengthens further the conceptual relation between RG and conformal methods.

A CFT can be fully characterized by providing the so-called \emph{CFT data},
which includes the scaling dimensions $\Delta_i$ of a set of operators known as primaries,
and the structure constants $C_{ijk}$ of their three point functions \cite{Osborn:1993cr, Erdmenger:1996yc, Fradkin:1996is}.
From the point of view of CFT, the scaling dimensions determine some of the most important properties of the scaling operators at criticality,
and in fact can be related to the critical exponents $\theta_i$ of an underlying second order phase transition,
while the structure constants provide further non-trivial information on the form of the correlators
of the theory. The CFT data can be used, in principle, to reconstruct the full model at, or close to, criticality.
In dimension greater than two
(see, for example, \cite{Osborn:1993cr, Erdmenger:1996yc, Fradkin:1996is}), 
however, since the symmetry group is finite dimensional, the use of analytical CFT methods is often not simple and
in fact most of the recent success of CFT applications comes from the numerical approach known as Conformal Bootstrap~\cite{ElShowk:2012ht}.
Up to now, RG methods have been almost always devoted to the computation of the RG spectrum
within a perturbative analysis in the $\epsilon$-expansion below the upper critical dimension of a given universality class.
The determination of the RG spectrum practically overlaps with the computation of the critical exponents
and thus of the scaling dimensions $\Delta_i$ of the underlying CFT.
It is thus natural to wonder to which extent the
RG can help the determination of the remaining CFT data: the structure constants $C_{ijk}$,
which have received far less attention in the RG literature.

The question which arises spontaneously is whether the approach of Cardy~\cite{Cardy:1996xt}
can be reversed and used to derive some of the OPE coefficients
once the RG flow of a model is known. 
In such a framework, since the beta functions are generally computed in a specific RG scheme,
one could expect that these RG based OPE coefficients might show some degree of scheme dependence.

The main purpose of this paper is to present an RG based approach, in a dimensionally regularized $\overline{\rm MS}$ scheme, 
to the computation of the OPE coefficients $\tilde{C}^k{}_{ij}$
related to the quadratic part of the Taylor expansion of the RG flow around a critical point.
We shall also pay attention to the transformation induced by general scheme changes among mass independent schemes,
and  infer some structure constants $C_{ijk}$ when scale invariance implies conformal invariance,
strengthening in this way the link between RG and CFT. 

We also show how the upgrade from {\it standard} perturbative RG to {\it functional} perturbative RG allows
for a more straightforward access to these quantities. 
After illustrating how to do this for two representative cases, the {\tt Ising} and (for the first time) the {\tt Lee-Yang} universality classes, respectively realized as unitary and non unitary theories, 
we also proceed to the construction of the beta functions for all the unitary (even) models relying heavily on the approach developed by 
O'Dwyer and Osborn \cite{ODwyer:2007brp}.
In general the use of the functional approach simplifies the computation of beta functions,
from which, in the vicinity of a fixed point, one can try to extract some of the (universal) CFT data  $\Delta_i$ and $C^k{}_{ij}$,
from linear and quadratic perturbations around the critical point, respectively.
This paper is concerned with fleshing out the main features of the functional approach and applying them
to the rich variety of critical theories which can be described with a single scalar field $\phi$.
The functional approach appears to be very powerful because the beta functionals are of a strikingly simple form,
and yet at the same time they describe the scaling behavior of classes of infinitely many composite operators.

In an effort to better understand the possible RG scheme dependence of the OPE coefficients $\tilde{C}^k{}_{ij}$
we carefully review their transformation properties.\footnote{In the context of conformal perturbation theory
this fact has already been discussed in~\cite{Gaberdiel:2008fn}.}
In our approach we can compute only the subset of OPE coefficients which
are ``massless'' at the upper critical dimension (the others being projected away by the dimensionally regularized scheme)
and therefore less sensitive to $\epsilon$-corrections induced by a change in the RG scheme. 
We show that our next-to-leading-order (NLO) computation gives these OPE coefficients at order $O(\epsilon)$ and
reproduces the structure constants previously obtained in a CFT framework~\cite{Codello:2017qek,Gliozzi:2016ysv}.
This fact, even if plausible, is in general not obvious because of the possible scheme dependence,
and we find it to be supported by the functional approach, which indeed constrains to some extent the possible choices of
coupling redefinitions that otherwise would be completely arbitrary.
All other ``massive'' $\tilde{C}^k{}_{ij}$ strongly depend on the computational scheme and vanish in dimensional regularization.
We observe that some OPE coefficients, including ``massive'' ones which would thus require a separate investigation,
can be or have already been obtained for several universality classes in the $\epsilon$-expansion in a CFT framework,
with either the CFT/Schwinger-Dyson Bootstrap \cite{Rychkov:2015naa, Basu:2015gpa, Raju:2015fza, Nii:2016lpa,Hasegawa:2016piv,Codello:2017qek}, 
the perturbative conformal block techniques~\cite{Gliozzi:2016ysv} or Mellin space methods \cite{Gopakumar:2016wkt}. Also large spin expansion techniques could be useful~\cite{Alday:2016njk}.

The first step in the functional perturbative RG approach is the computation of the beta functional $\beta_V$ of the effective potential $V(\phi)$,
which generates the beta functions of the couplings of all the local operators $\phi^k$. 
This can often be used to verify our results by checking them against the renormalization of the relevant operators.
Next comes the inclusion of the beta functional $\beta_Z$ of a field dependent wavefunction $Z(\phi)$,
which generates the flow of the couplings corresponding to operators of the form $\phi^k (\partial \phi)^2$ and,
through its boundary conditions, allows also for the determination of the anomalous dimension $\eta$.
Higher-derivative operators can be added on top of the aforementioned ones
following a construction based on the derivative expansion, which treats operator mixing in a systematic way, a topic that will be discussed here later on.

The content of the paper is as follows:
In Sect.~\ref{gen_RG_CFT} we show in general how to use the RG to compute both the spectrum of scaling dimensions and the coefficients of the OPE.
We discuss in general the possible scheme dependence  by studying their behavior under arbitrary changes of parametrization of the space of all couplings,
and we illustrate our methods by considering the RG flow of the {\tt Ising}~\cite{zinn-justin,Kleinert:2001ax} universality class
as an example.\footnote{We will pursue the convention of \cite{Codello:2017qek} and denote universality classes with the {\tt Typewriter} font.
This is meant to avoid any confusion between the universality classes and the models realizing them at criticality.
For example, the {\tt Ising} universality class and the Ising spin $\pm 1$ on a lattice are generally distinguished,
with the latter behaving according to the former only at the critical temperature and at zero magnetic field.}
In Sect.~\ref{section-tutorials} we promote the standard perturbative RG to functional perturbative RG 
and illustrate the procedure by applying it to the {\tt Ising} and 
{\tt Lee-Yang}~\cite{Fisher:1978pf, Cardy:1985yy, deAlcantaraBonfim:1980pe, Macfarlane:1974vp, Gracey:2015tta, An:2016lni, Zambelli:2016cbw} universality classes.
Using the beta functionals for the effective potentials in these two examples, we give general formulas for both the spectrum and the structure constants of the underlying CFTs,
and use them to highlight the main novelties of the approach.
In Sect.~\ref{sec de} we describe how to systematically improve the functional approach, to include arbitrary higher derivative operators, and how to generally deal with operator mixing.
In Sect.~\ref{multi-critical}  we present 
an application of this framework
to  the study of the general multicritical $\phi^{2n}$ universality class~\cite{Itzykson:1989sx, ODwyer:2007brp}.
Finally, in Sect.~\ref{sect-conclusions} we draw some conclusions and discuss the most important prospects of this approach.

In appendix~\ref{osborn} we review the perturbative computations which are necessary to obtain the beta functionals used in 
Sect.~\ref{multi-critical}.
In appendix~\ref{genscaling} we show how to use the functional approach to prove some simple scaling relations between critical exponents generally known as shadow relations.
In appendix~\ref{relFRG} we spell out some intriguing relation between the perturbative and 
non-perturbative functional RG approaches~\cite{Wegner:1972ih, Polchinski:1983gv, Wetterich:1992yh, Morris:1994ie}.

\section{Spectrum and OPE coefficients from RG analysis}\label{gen_RG_CFT}

The primary goal of the RG analysis is the study of universality classes and the determination of their quantitative properties, 
i.e.\ the CFT data (when the two are related). 
This data is the union of the spectrum (the set of scaling dimensions $\Delta_i$ of composite operators) and the set of structure constants 
(in a CFT these are in one-to-one correspondence with the OPE coefficients $C^k{}_{ij}$ of primary operators).

The aim of this section is to introduce a computational scheme which shows how CFT data is (partially) encoded in the beta functions 
describing the RG flow in proximity of a fixed point. We start by describing the picture recalling a picture inspired by Cardy~\cite{Cardy:1996xt}, 
which was originally defined in a short distance regularized scheme, and considering a generic basis of operators in which possible mixing effects are present.
Then we present, in a dimensionally regularized scheme, a simple discussion of the {\tt Ising} universality class 
to provide an example of an RG determination of OPE coefficients in the $\epsilon$-expansion, 
which will also motivate the subsequent discussion of the transformation properties of the $\tilde{C}^k{}_{ij}$. 
This discussion will make clear which subset of OPE coefficients can actually appear in the beta functions, 
finally explaining which part of the CFT data is directly accessible by our RG methods.

\subsection{General analysis}\label{gen_an}

We begin our analysis by considering a general (renormalized) action in $d$ dimensions,
\begin{equation}
S=\sum_i \, \mu^{d-\Delta_i} g^i\int {\rm d}^d x  \,\Phi_i(x) \,,
\label{generalS}
\end{equation}
describing an arbitrary point in theory space.
The choice of a basis set of operators $\Phi_i$ allows the introduction of coordinates, i.e.\ the corresponding (dimensionless) couplings $g^i$.  
The scaling dimensions $\Delta_i$ of the (composite) operators, as we will see in a moment, take precise values only in the vicinity of a fixed point of the RG flow.  
All information regarding the flow can be extracted from the set of (dimensionless) beta functions $$\beta^i=\mu \frac{{\rm d} g^i}{{\rm d} \mu}\,,$$ 
which are in principle fully computable once a given scheme  is precisely defined. 
A fixed point of the RG flow is the point $g_*^i$ in the space of couplings for which the theory is scale invariant
\begin{equation}
 \begin{split}
 \beta^i(g_*) = 0\,.
 \end{split}
\end{equation}
In the neighborhood of a fixed point it is convenient to characterize the flow by Taylor expanding the beta functions.
If $\delta g^i$ parametrizes the deviation from the fixed point ($g^i=g^i_*+\delta g^i$), we have
\begin{equation}
 \begin{split}
 \beta^k(g_*+\delta g)  = \sum_i M^k{}_{i}\, \delta g^i+ \sum_{i,j} N^k{}_{ij} \,\delta g^i \, \delta g^j +O(\delta g ^3)\,,
 \end{split}
\end{equation}
where at the linear level we defined the stability matrix
\begin{equation}
 \begin{split}
 M^i{}_{j} &\equiv \left.\frac{\partial\beta^i}{\partial g^j}\right|_{*} \\
 \end{split} \label{m}
\end{equation}
and at the quadratic level we defined the tensor
\begin{equation}
 \begin{split}
 N^i{}_{jk} & \equiv \frac{1}{2} \left.\frac{\partial^2\beta^i}{\partial g^j\partial g^k}\right|_{*}  \,,
 \end{split} \label{n}
\end{equation}
which is symmetric in the last two (lower) indices.

Each scale invariant point of the RG flow is in one to one correspondence with a universality class and, under mild conditions that we assume, a related CFT.
The spectrum of the theory at criticality is given by the eigendeformations of $M^i{}_{j}$ with the corresponding eigenvalues being (the negative of) the critical exponents $\theta_a$.
We will only be concerned with cases in which either the matrix $M^i{}_{j}$ is already diagonal, 
or its left and right spectra coincide (meaning that the spectrum is unique and unambiguous). 
It is convenient to introduce the rotated basis $\lambda^a = \sum_i {\cal S}^a{}_i \,\delta g^i$ 
which diagonalizes $M^i{}_{j}$ (through the linear transformation ${\cal S}^a{}_i \equiv  \partial \lambda^a / \partial  \delta g^i \big|_* $)
\begin{equation}
 \begin{split}
 \sum_{i,j} {\cal S}^a{}_i \, M^i{}_{j} \,({\cal S}^{-1})^j{}_b & = -\theta_{a} \delta^a{}_{b}\,.
 \end{split}
\end{equation}
Critical exponents allow for a precise definition of the scaling dimensions of the operators through the relation $\theta_i = d - \Delta_i$.
%
Let us introduce the ``canonical'' dimensions $D_i$ of the couplings, and parametrize the deviations of the critical exponents
from the canonical scaling through the anomalous dimensions $\tilde \gamma_i$ as
%
\begin{equation}
 \theta_i = d- D_i - \tilde \gamma_i \,.
\end{equation}
Here and in the following we adopt a tilde to distinguish RG quantities from CFT ones.
The notion of canonical dimension is in principle arbitrary, but in real-world applications
it is generally borrowed from the scaling of the {\tt Gaussian} critical theory.

This expression is, strictly speaking, valid only for primary operators; for descendants there is a subtlety that we will discuss later.
The matrix $({\cal S}^{-1})^i{}_a$ also returns the basis of scaling operators of the theory at criticality, ${\cal O}_a=\sum_i ({\cal S}^{-1})^i{}_a\, \Phi_i$,
so that we can rewrite the action as a fixed point action (i.e.\ CFT action) plus deformations 
\begin{equation}
S=S_* + \sum_a \, \mu^{\theta_a} \lambda^a \int {\rm d}^d x  \,{\cal O}_a(x) +O(\lambda^2)\,.
\label{CFTPT}
\end{equation}
Deformations are relevant, marginal or irrelevant depending on the value of the related critical exponent (respectively positive, zero or negative).
In the diagonal basis also the tensor $N^i{}_{jk}$ have a direct physical meaning,
since after the diagonalizing transformation it becomes a quantity related to
the (symmetrized) OPE coefficients\footnote{Note that the overall normalization 
of the OPE coefficients is not fixed: a rescaling of the couplings $\lambda^a \to \alpha_a \lambda^a$ implies $\tilde C^a{}_{bc} \to \frac{\alpha_b \alpha_c}{\alpha_a} \tilde C^a{}_{bc}$.}
\begin{equation}
 \begin{split}
\tilde C^a{}_{bc} &= \sum_{i,j,k} {\cal S}^a{}_i \, N^i{}_{jk} \, ({\cal S}^{-1})^j{}_b \, ({\cal S}^{-1})^k{}_c \,,
 \label{opeC}
 \end{split}
\end{equation}
It will become clear in the practical examples that will follow this subsection that at $d=d_c$ the $\tilde C^a{}_{bc}$ are the OPE coefficients of the underlying
{\tt Gaussian} CFT and that all $O(\epsilon)$ corrections agree with CFT results for all available comparisons,
despite the general inhomogeneous transformations of these coefficients under general scheme changes as discussed in subsection~\ref{transformation}.
For these reasons we make the educated guess that the quantities in \eqref{opeC} are the $\overline{\rm MS}$ OPE coefficients 
since they have been computed using $\overline{\rm MS}$ methods.
The relation among the standard perturbative $\overline{\rm MS}$ OPE coefficients and quadratic coefficients in the beta functions is an interesting subject, 
which is nevertheless beyond the scope of this work and is left for future investigations.


The beta functions can now be written as
\begin{equation}
 \begin{split}
\beta^a  = -(d-\Delta_{a}) \lambda^a + \sum_{b,c} \tilde C^a{}_{bc}\, \lambda^b \lambda^c +O(\lambda^3)\,.
\label{betaFP}
 \end{split}
\end{equation}
This formula is the familiar expression for beta functions in CFT perturbation theory (see, for example, \cite{Cardy:1996xt}) and provides a link between RG and CFT.
Generalizations of this result beyond the leading order are considerably less simple than what we presented here \cite{Gaberdiel:2008fn}.

In CFT one uses the OPE\footnote{These OPE coefficients are related to those entering the beta functions by a factor $S_d/2$ (see~\cite{Cardy:1996xt}).}  
\begin{equation}
\begin{split}
\left\langle {\cal O}_a(x) \,{\cal O}_b(y) \cdots \right\rangle = \sum_c 
\frac{1}{\left|x-y\right|^{\Delta_a+\Delta_b-\Delta_c}} \,C^c{}_{ab} \left\langle {\cal O}_c(x) \cdots \right\rangle
 \end{split}
\end{equation}
to renormalize a perturbative expansion of the form \eqref{CFTPT} in which the CFT is described by the action $S_*$
and deformations are parametrized by the couplings $\lambda^a$.\footnote{The careful reader must have noticed that our determination of the $C^a{}_{bc}$ is symmetrized in the lower two indices,
but it is more than enough to reconstruct the fully symmetric structure constants $C_{abc}$.}
In the RG framework, conversely, the knowledge of the beta functions could permit (in principle) the extraction of the conformal data
directly from \eqref{betaFP}.
The rest of this paper is essentially devoted to a detailed exploration of this link, first within a simple example in the next subsection and then,
after a short discussion of the scheme dependences of the OPE coefficients, within a functional generalization of standard perturbation theory $\epsilon$-expansion.

\subsection{Example: {\tt Ising} universality class}\label{ising0}

It is useful at this point to consider an explicit example to introduce our approach, the {\tt Ising} universality class in $d=4-\epsilon$~\cite{zinn-justin,Kleinert:2001ax}. 
Perturbation theory forces us to restrict to deformations around the Gaussian fixed point, 
the simplest of which are power like non derivative operators $\Phi_i = \phi^i$ parametrized by the dimensionless couplings $g_i$, as will be shown in the next section.

One can obtain the (two loop) NLO beta functions for relevant and marginal 
deformations\footnote{Here and in other sections with explicit examples we lower the vector indices of the beta functions and the couplings to avoid any confusion with power exponents.}
\begin{equation}
 \begin{split}
  \beta_1 &= -\!\left(3-\frac{\epsilon}{2}\right)g_1 +12\, g_2g_3 -108\,g_3^3 -288\,g_2g_3g_4+48\,g_1g_4^2 \\
  \beta_2 &= -2\,g_2 +24 \,g_4g_2 +18\,g_3^2 -1080\,g_3^2g_4 -480\,g_2g_4^2 \\
  \beta_3 &= -\!\left(1+\frac{\epsilon}{2}\right)g_3 +72\,g_4g_3 -3312\,g_3g_4^2\\
  \beta_4 &= -\epsilon g_4 + 72\,g_4^2 -3264\,g_4^3
  \label{betaising1234}
 \end{split}
\end{equation}
and the anomalous dimension $\eta = 96g_4^2$. 
Note that the coefficients of the one loop leading-order (LO) quadratic terms in the couplings are directly related to the {\tt Gaussian} 
OPE coefficients, which by construction coincide with the mean field OPE coefficients
of the {\tt Ising} universality class (see also~\cite{Hogervorst:2015akt}).

The fixed point is characterized by 
$g_4^*=\frac{\epsilon}{72}+\frac{17\epsilon^2}{1944}+O(\epsilon^3)$
and $g_1^*=g_2^*=g_3^*=0$.
Around this fixed point one therefore expands in powers of deformations 
(with $\lambda_i = g_i$ for $i=1,2,3$ and $\lambda_4 = g_4-g_4^*$),
and the beta functions become
\begin{equation}
 \begin{split}
 \beta_1 &=
  -\left(3-\frac{\epsilon}{2}-\frac{\epsilon^{2}}{108}\right)\!\lambda_{1}
  +12\left(1-\frac{\epsilon}{3}\right)\!\lambda_{2}\lambda_{3}
  +\frac{4}{3}\epsilon \,\lambda_{1}\lambda_{4}+\dots
  \\
 \beta_2 &=
  -\left(2-\frac{\epsilon}{3}-\frac{19\epsilon^{2}}{162}\right)\!\lambda_{2}
  +24 \left(1-\frac{5}{9}\epsilon\right)\!
  \lambda_{2}\lambda_{4}
  +18 \left(1-\frac{5}{6}\epsilon\right)\!\lambda_{3}^{2}
  +\dots\\
 \beta_3 &=
  -\left(1-\frac{\epsilon}{2}+\frac{\epsilon^{2}}{108}\right)\!\lambda_{3}
  +72 \left(1-\frac{23}{18}\epsilon\right)\!
  \lambda_{3}\lambda_{4}
  +\dots\\
 \beta_4 &=
  -\left(-\epsilon+\frac{17\epsilon^{2}}{27}\right)\!\lambda_{4}
  +72 \left(1-\frac{17}{9}\epsilon\right)\!
  \lambda_{4}^{2}+\dots
  \label{betafp}
 \end{split}
\end{equation}
One should keep in mind that the NLO coefficients of quadratic terms involving $\lambda_4$ can be affected by diagonalization. 
From the above relations one can immediately read off the critical exponents $\theta_1,\theta_2,\theta_3,\theta_4$ as minus the coefficients of the linear terms.
Note that the scaling relation $\theta_1+\theta_3 = d$ discussed in appendix~\ref{genscaling} is indeed satisfied.
We will see that these couplings will not be subject to any further mixing and thus these are the complete $\epsilon$-series to the exhibited order for the critical exponents 
and the OPE coefficients.
We have limited the $\epsilon$-series for the OPE coefficients to linear order since the $O(\epsilon^2)$ terms are incomplete, 
receiving contributions from next-to-next-to-leading-order (NNLO) beta functions.

From the eigenvalues we can extract the coupling (RG) anomalous dimensions $\tilde{\gamma}_i$ through the relations
\begin{equation}
\theta_i = d-i\left(\frac{d-2}{2}\right) - \tilde\gamma_i
\label{theta}
\end{equation}
and $\eta = 2 \tilde \gamma_1$. The scaling dimensions of the composite operators are instead
\begin{equation}
\Delta_i = i\left(\frac{d-2}{2}\right) + \gamma_i
\label{Delta}
\end{equation}
and define the (CFT) anomalous dimensions $\gamma_i$.
The difference between the $\tilde \gamma_i$ and $\gamma_i$ appears only when the related operators are descendant, 
in this case when $i=3$ for which $\gamma_3=\tilde\gamma_3+\eta$.
We will postpone the discussion of this fact to the appendix~\ref{genscaling}.
The explicit expressions for the first anomalous dimensions are well known
\begin{equation}
\tilde \gamma_1 = \frac{\epsilon^2}{108} \qquad\tilde \gamma_2 = \frac{\epsilon}{3} +\frac{19\epsilon^2}{162}
 \qquad\tilde \gamma_3 = \epsilon -\frac{\epsilon^2}{108} \qquad\tilde \gamma_4 = 2\epsilon -\frac{17\epsilon^2}{27}\,. \nonumber
\end{equation}
From \eqref{betafp} it is equivalently easy to read off the OPE coefficients
(which on the non-diagonal entries are half the value of the coefficients in the beta functions)
\begin{equation} \label{Cising}
\tilde C^1{}_{23}=6-2\epsilon \qquad \tilde C^{1}{}_{14} = \frac{2}{3}\epsilon \qquad  \tilde C^{2}{}_{33}= 18-15\epsilon 
\end{equation}
We note that the OPE coefficient $\tilde C^{1}{}_{14}$ is in perfect agreement with that found in~\cite{Codello:2017qek} 
using CFT methods, while we did not find any result in the literature for the other two coefficients to compare to.
An explanation of why this agreement is expected will be given in Sect.~\ref{ising1}.
It is also important to stress that we ensured the agreement by choosing the same normalization of~\cite{Codello:2017qek},
that is by fixing the coefficients of the two point functions.

\subsection{Transformation properties}\label{transformation}

In general different regularization and renormalization procedures may result into non trivial relations among the renormalized couplings.
These relations go under the name of scheme transformations, and they are exemplified through maps among the couplings of the two schemes that can be
highly non-linear~\cite{Weinberg:1996kr}.
Whenever the scheme transformations are computed between two mass independent schemes (such as, for example, $\overline{\rm MS}$ 
and lattice's\footnote{But in practice all lattice implementations can be considered massive schemes.}) 
these relations might have a simpler form, but we will find that it is very useful to consider them in their most general form.
Let
\begin{equation} 
\begin{split}
\bar g^{i} &
= \bar g^i(g)
\end{split}
\end{equation}
be the general invertible, possibly non-linear, transformation between the set of couplings $g^i$ and $\bar g^i$.
Under such a change of ``coordinates'' the beta functions transform as vectors\footnote{The suummation convention is understood in this subsection.}
\begin{equation} 
\begin{split}
\bar \beta^i(\bar g) &
= \frac{\partial \bar g^i}{\partial  g^j}\, \beta^j(g)\,.
\end{split}
\end{equation}
Now we turn our attention to the fixed point quantities, therefore in the following it is understood that the transformations will be evaluated at a fixed point.
The stability matrix transforms as
\begin{equation} 
\begin{split} 
\bar M^i{}_{j} &
= \frac{\partial \bar g^i}{\partial g^l}\, M^l{}_{k} \, \frac{\partial g^k}{\partial \bar g^j}\,.
\end{split}
\end{equation}
Since the derivatives are evaluated at the fixed point, the stability matrices of the two set of couplings are related by a similarity transformation. Therefore it is trivial to prove that the spectrum is invariant, meaning that it does not depend on the parametrization
\begin{equation} 
\begin{split} 
\bar \theta_a &
= \theta_a\,,
\end{split}
\end{equation}
as one would naively expect for a physical quantity.

Things become less trivial when considering the matrix encoding the second order of the Taylor expansion at the fixed point.
A direct computation shows
\begin{equation} 
\begin{split}
\bar N^i{}_{jk} = &
\frac{\partial \bar g^i}{\partial g^c} \Bigl\{
N^c{}_{ab} + \frac{1}{2} M^c{}_{d} \frac{\partial^2 g^d}{\partial \bar g^l\partial \bar g^m}\frac{\partial \bar g^l}{\partial g^a}\frac{\partial \bar g^m}{\partial g^b}
- \frac{1}{2} M^d{}_{a} \frac{\partial^2 g^c}{\partial \bar g^l\partial \bar g^m}\frac{\partial \bar g^l}{\partial g^b}\frac{\partial \bar g^m}{\partial g^d}
\\&
\qquad- \frac{1}{2} M^d{}_{b} \frac{\partial^2 g^c}{\partial \bar g^l\partial \bar g^m}\frac{\partial \bar g^l}{\partial g^a}\frac{\partial \bar g^m}{\partial g^d}
\Bigr\} \frac{\partial g^a}{\partial \bar g^k} \frac{\partial g^b}{\partial \bar g^j}\,.
\end{split}
\end{equation}
To simplify this expression it is convenient to assume that the couplings $g_k$ have already been chosen to diagonalize the stability matrix with a linear transformation, so that on the right hand side there will be the structure constants
\begin{equation} 
\begin{split}
\bar N^i{}_{jk} &= \frac{\partial \bar g^i}{\partial g^c} \left\{ \tilde C^c{}_{ab} + \frac{1}{2} (\theta_c - \theta_a - \theta_b) \frac{\partial^2 g^c}{\partial \bar g^l\partial \bar g^m}\frac{\partial \bar g^l}{\partial g^a}\frac{\partial \bar g^m}{\partial g^b}\right\} \frac{\partial g^a}{\partial \bar g^k} \frac{\partial g^b}{\partial \bar g^j}\,.
\end{split}
\end{equation}
Now it is necessary to move to the basis of couplings $\bar g^i$ in which $\bar M^i{}_{j}$ is diagonal, so that the structure constants appear on both sides.
We finally find
\begin{equation} 
\begin{split}
\bar{ \tilde C}^c{}_{ab} =\tilde C^c{}_{ab} + \frac{1}{2}\, (\theta_c - \theta_a - \theta_b)\, \frac{\partial^2 g^c}{\partial \bar g^l\partial \bar g^m}\frac{\partial \bar g^l}{\partial g^a}\frac{\partial \bar g^m}{\partial g^b}\,,
\end{split}
\label{opeCtransform}
\end{equation}
which implies that the set of matrices $\tilde C^c{}_{ab}$ has a transformation law that is not homogeneous and therefore is reminiscent of the one of a connection in the space of couplings \cite{Kutasov:1988xb,Lassig:1989tc}. 

In the context of conformal perturbation theory one can find a similar result in~\cite{Gaberdiel:2008fn}, in which the analysis includes cubic terms but is limited to a diagonal stability matrix
because conformal perturbation theory adopts by construction the basis of scaling operators.

At this point few comments on the transformations of $\tilde C^c{}_{ab}$ are in order:
\begin{itemize}

\item 
It is evident from \eqref{opeCtransform} that $\tilde C^c{}_{ab}$ can be independent of parametrization for a very special sum condition among the scaling dimensions of the couplings or when the Hessian at the fixed point is zero. The latter case could be realized for a specific family of scheme transformations, while the former condition can be realized exactly only in $d=d_c$, which corresponds to $\epsilon =0$, that is,
\begin{equation} 
\left(\theta_c - \theta_a - \theta_b\right)_{\epsilon=0}=0\,.
\label{universal}
\end{equation}

\item We will observe in all practical examples that the coefficients $\tilde C^c{}_{ab}$ for which the condition in Eq.~\eqref{universal} holds
are the ones that are accessible via dimensional regularization. We dub them ``massless",
as opposed to the ``massive" ones that do not satisfy the above condition and are zero in dimensional regularization.
Moreover, these ``massless'' OPE coefficients at the critical dimension are insensitive to changes of RG scheme
and can be computed unambiguously with RG methods.

\item In perturbation theory $\epsilon$-expansion one can obtain $\epsilon$-series only for the ``massless'' OPE coefficients. 
In particular in $d=d_c -\epsilon$ one generally has $\theta_c - \theta_a - \theta_b=O(\epsilon)$ and 
thus only the $O(\epsilon)$ terms can be scheme independent if the Hessian is at least $O(\epsilon)$.
This allows for crucial comparisons and cross-checks with other theoretical approaches like CFT (as for the $\tilde C^1{}_{14}$ of our previous example).
In fact we will see that all the $\overline{\rm MS}$ leading corrections for the multi-critical models we can compare with CFT are in perfect agreement.
While this agreement can be explained at the level of beta functions, by explicitly constructing the most general map between to orthogonal massless scheme that also preserves the $\epsilon$-expansion, the explanation is more transparent when discussed in functional terms in the next section.

\end{itemize}


Finally one should remark that once the beta functions are extracted in some scheme, one might also envisage geometrical methods to extract quantities which depend on the universal scheme independent OPE coefficients to overcome the above limitations.
A step in this direction has been made for functional-type flows in Ref.~\cite{Lizana:2017sjz} in the context of the Polchinski RG equation.
In this work, the authors define normal coordinates in the space of couplings which have both geometrical meaning and definite scaling transformations.
In relation with the transformation \eqref{opeCtransform}, one can follow \cite{Lizana:2017sjz} and argue that all ``massive'' coefficients
can be eliminated by an opportune transformation of the couplings and hence there exists a scheme, or rather a family of schemes,
whose only coefficients are the scheme independent ones. This family can be appropriately named ``family of minimal subtraction schemes'' having the $\overline{\rm MS}$ scheme as its most famous representative.
We hope to address further these topics in future investigations.

\section{Functional perturbative RG: a first look} 
\label{section-tutorials}

The previous example on the {\tt Ising} universality class, which was dealing with the study of the ${\cal O}_k=\phi^k$ deformations of the Gaussian fixed point,
can be analyzed more conveniently if we work directly with the generating function of such operators which is the local potential $V(\phi)$.
We thus consider the action
\begin{equation}
S=\int {\rm d}^d x \left\{\frac{1}{2}(\partial\phi)^2 + V(\phi) \right\}\,,
\label{LPA}
\end{equation}
and study the perturbative RG  flow it generates. In particular, it turns out to be a smart move to perform background field computations
of loop diagrams in which the field $\phi$ is set to a constant and leave the form of the potential completely general
so that we can extract the beta functional $\beta_V$ for the {\it whole} potential just by looking at the vacuum renormalization.
This way of thinking has at least a two fold advantage: it simplifies computations (since we just need to compute the vacuum renormalization)
and gives direct access to the full system of beta functions for the couplings of the operators ${\cal O}_k=\phi^k$
(since $\beta_V$ serves as a {\it generating function for the beta functions}). From the knowledge of the beta functions
we can then follow the steps outlined in the previous section and compute both the spectrum and the OPE coefficients in the $\overline{\rm MS}$ scheme.

The action \eqref{LPA} not only renormalizes the potential, but also induces the flow $\beta_Z$ of a field-dependent wavefunction functional that we will denote $Z(\phi)$.
The flow generates the beta functions of the couplings of the operators of the form $\phi^k (\partial\phi)^2$
and, moreover, fixes the anomalous dimension $\eta$. More generally, all higher derivative operators have an approximate flow
induced solely by the potential, i.e.\ have a beta functional whose r.h.s.\ contains only $V(\phi)$ and its derivatives.
We will call local potential approximation (LPA) the truncation for which all the RG flow is generated by the potential alone. 
Clearly, the full RG flow will involve the presence of other functionals, such as $Z(\phi)$ and higher, on the r.h.s.\ of the beta functionals.
According to that the computational scheme can be systematically improved in a derivative expansion approach, as will be discussed in section~\ref{sec de}.

In this section we will study, as a tutorial example, the {\tt Ising} and {\tt Lee-Yang} universality classes within the LPA, while 
a first example of functional flow beyond this approximation will be presented in section~\ref{multi-critical}, in which we show that terms containing $Z(\phi)$
on the r.h.s.\ of the beta functional $\beta_Z$ become important to describe mixing effects when marginal or irrelevant operators are investigated in the  {\tt Ising} and  multicritical universality classes. 

For any given theory and within a functional perturbative approach in a dimensionally regularized scheme, e.g.\ $\overline{\rm MS}$, such beta functionals
can be written as polynomials for which each monomial is a product of derivatives of various orders of the generating functions $V$, $Z$, \dots \
and in particular each non trivial loop order in perturbation theory gives rise to a subset of monomials in the beta functionals \cite{ODwyer:2007brp}. 
Let us just stress that for a given theory the monomials which can appear in the beta functional are very constrained and only their coefficients demand a real loop computation, which in turn can be
done in very specific and simple ways.

Another point to highlight is that, depending on the specific theory, contributions denoted as LO (or NLO or higher) appear at different number of loops, generally bigger than one
({\tt Ising} and {\tt Lee-Yang} are special in this respect since the LO terms are obtained at one loop and NLO at two loops for both theories).
We shall see this explicitly in section~\ref{multi-critical} in which we study the whole family of multicritical $\phi^{2n}$ universality classes: indeed the number of loops required to obtain the LO contribution
depends on the critical dimension $d_c$ of the theory (which determines  the superficial degree of divergence of a diagrams generated by the perturbative expansion).
One sees that, since each member of this family of models has  $d_c=\frac{2n}{n-1}$ (for $n>1$), the leading order contribution appears at $(n-1)$-loops and the NLO at $2(n-1)$-loops.
Let us also mention the fact that the LO and NLO order contributions are universal,
i.e.\ independent of the specific RG scheme (as can be easily seen by projecting the beta functionals
on the beta function of the respective critical coupling which we already know has LO and NLO universal coefficients~\cite{Weinberg:1996kr}).

It is also convenient to make the standard shift to dimensionless variables (in units of the scale $\mu$) directly at the functional level.
Once the beta functional of the dimensionful potential is found, the scaling properties are investigated by defining the dimensionless potential
\begin{equation}
v(\varphi)=\mu^{-d}V(\varphi \mu^{d/2-1} Z_0^{-1/2})
\label{dimless}
\end{equation}
where $Z_0$ is the field strength renormalization, which enters in the definition of the dimensionless field $\varphi=\mu^{1-d/2} Z_0^{1/2} \phi$.
Its beta functional is then
\begin{equation}
 \begin{split}
\beta_v &= -d v(\varphi) + \frac{1}{2}(d-2+\eta)\varphi \ v^{(1)}(\varphi)+ \mu^{-d} \beta_V\,,\\
 \end{split}
 \label{betav}
\end{equation}
for which we introduced an anomalous dimension $\eta=- \mu\partial_\mu \log Z_0$, which will be discussed in detail soon.

The potential is a local function of the dimensionless field $\varphi$ and can be parametrized in terms of the dimensionless couplings $g_k$ as
\begin{equation}
 \begin{split}
  v(\varphi) &= \sum_{k \geq 0}  g_k \varphi^k\,.
 \end{split}
 \label{eq:potential_couplings}
\end{equation}
The beta functional is then used to obtain the couplings' beta functions through the straightforward definition
\begin{equation}
 \begin{split}
  \beta_v &=
  \sum_{k \geq 0}\beta_k \varphi^k\,.
  \label{bvcoup}
 \end{split}
\end{equation}
One then inserts \eqref{eq:potential_couplings} and \eqref{bvcoup} 
on the r.h.s.\ and l.h.s.\ of \eqref{betav}, respectively, and equates powers of the field on both sides to obtain the general beta function system. 

The dimensionless wavefunction is similarly defined as
$$z(\varphi)=Z_0^{-1} Z(\varphi \mu^{d/2-1} Z_0^{-1/2})$$
and its dimensionless flow is
\begin{equation}
 \begin{split}
\beta_z &= \eta z(\varphi) + \frac{1}{2}(d-2+\eta)\varphi \ z^{(1)}(\varphi)+Z_0^{-1} \beta_Z\,.\\
 \end{split}
  \label{betaz}
\end{equation}
This new beta functional has two main purposes. On the one hand by enforcing the condition $z(0)=1$
we can use it to determine $\eta$ as
\begin{equation}
 \begin{split}
  \eta & = - \mu\partial_\mu \log Z_0= - Z_0^{-1} \beta_Z(0)\,.
 \label{eta}
 \end{split}
\end{equation}
On the other hand, later in section~\ref{multi-critical} we will use \eqref{betaz} to generate the beta functions of the dimensionless couplings of the operators of the form $\varphi^k (\partial \varphi)^2$.
A detailed discussion of the invariance of the systems of beta functionals in the LPA under reparametrizations of $z(0)$, and of its importance in the determination of $\eta$ can be found in \cite{Osborn:2009vs}.

\subsection{{\tt Ising} universality class in LPA}\label{ising1}

The {\tt Ising} universality class has upper critical dimension $d_c=4$, and the LPA beta functionals for the dimensionful potential at NLO, and wavefunction at LO, are
\begin{equation}
  \beta_V = \frac{1}{2}\frac{(V^{(2)})^2}{(4\pi)^2}-\frac{1}{2}\frac{V^{(2)}(V^{(3)})^2}{(4\pi)^4}
 \qquad \qquad
  \beta_Z = -\frac{1}{6}\frac{(V^{(4)})^2}{(4\pi)^4}\,.
\end{equation}
The functional form of these beta functionals can be argued on dimensional grounds. 
Only the explicit determination of the three universal coefficients demand a loop computation, but 
for a well studied universality class such as {\tt Ising} these coefficients can be obtained by matching with known beta functions of the $\phi^4$ critical coupling.

In turn, this simple observation shows that these coefficients are scheme independent by the standard text book argument
that LO and NLO beta functions and anomalous dimension coefficients are so. Thus all the results of the present section,
and in particular the form of the beta functions around the fixed point are a functions of these universal numbers.
In particular this implies that the order $\epsilon$ and $\epsilon^2$ contributions to the spectrum (which is universal)  are scheme independent, and that the order $\epsilon$ corrections to the OPE coefficients are also scheme independent
even if the $\tilde C^i{}_{jk}$ themselves are not universal. As promised this is a simple way to understand why the Hessian for the "dimless" OPEs is of order at least $O(\epsilon)$ \cite{CSVZ5}.


After a simple rescaling $v \to (4\pi)^2\, v$, the beta functionals for the dimensionless potential are the following
\begin{equation}
 \begin{split}
  \beta_v &= 
  -4 v + \varphi v^{(1)}
  +\epsilon \left(v -\frac{1}{2}\varphi  v^{(1)}\!\right)
  +\frac{1}{2}\eta\varphi v^{(1)}
  +\frac{1}{2}(v^{(2)})^2-\frac{1}{2}v^{(2)}(v^{(3)})^2
  \\
  \beta_z &= 
  \eta z + \varphi z^{(1)}
  -\frac{\epsilon}{2}\varphi z^{(1)}
  +\frac{1}{2}\eta\varphi z^{(1)}
  -\frac{1}{6}(v^{(4)})^2\,.
  \label{LPAising}
 \end{split}
\end{equation}
Expanding the potential as in Eq.~\eqref{eq:potential_couplings} discussed before,
allows the generation of the coupling's beta functions 
\begin{equation}
 \begin{split}
  \beta_k =&
  -\!\left(4-k-\Big(1-\frac{k}{2}\Big)\epsilon\right)g_k
  +48  k  \,g_k \,g_4^2
  \\&
  +\frac{1}{2} \sum_{i=2}^{k+2} i(i-1) (i-k-4)(i-k-3) \,g_i\, g_{4-i+k}
  \\&
 + \frac{1}{6} \sum_{i=2}^{k+7}\,\,\, \sum_{j=2}^{k+7-i} \, i(i-1)  j(j-1)  (i+j-k-8) (i+j-k-7) \times
  \\&\qquad \times \left(i^2+ij+j^2  -(i+j)(k+8)+4 (k+5)\right) g_i\, g_j\,g_{8-i-j+k}
  \,. \label{betaising}
 \end{split}
\end{equation}
The four-coupling system \eqref{betaising1234} studied in the previous section is straightforwardly obtained by truncating \eqref{betaising} to $k=1,2,3,4$.
Using \eqref{eta} we can immediately obtain the anomalous dimension
\begin{equation}
\eta = \frac{1}{6} (v^{(4)}(0))^2 = 96 g_4^2\,.
\label{etaising}
\end{equation}
In dimensional regularization the fixed point is very simple
\begin{equation}
g_k^* = g\, \delta_{k,4} \qquad\qquad
 g = \frac{\epsilon}{72}+\frac{17\epsilon^2}{1944}
\end{equation}
%
and highlights the prominent role of the critical coupling $g_4$.
By expanding around the fixed point it is straightforward to obtain 
the following general form for the spectrum (in terms of the critical coupling)
%
\begin{equation}
  \theta_i
  = 4-i-\left(1-\frac{i}{2}\right)\epsilon-\frac{1}{2}i(i-1)g+\frac{1}{12}i(6i^2-12i+5)g^2-\frac{2}{3}g^2 \delta_{i,4}\,.
  \label{thetaisingi}
\end{equation}
%
Using \eqref{theta} we immediately deduce the anomalous dimensions of the composite operators $\phi^i$ (in terms of $\epsilon$)
\begin{equation}
\tilde \gamma_i = \frac{1}{6}i(i-1)\epsilon-\frac{1}{324}i(18i^2-70i+49)\epsilon^2+\frac{2}{27}\epsilon^2\delta_{i,4}\,.
\label{gammaisingi}
\end{equation}
For $i=1,2,3,4$ this expression reproduces those of the example in the previous section.
The reader will notice the appearance of a contribution in \eqref{thetaisingi} and in \eqref{gammaisingi} proportional to the Kronecker delta $\delta_{i,4}$
because the anomalous dimension in \eqref{LPAising} 
is a function of the critical coupling as given in \eqref{etaising}~\cite{zinn-justin,Kleinert:2001ax}.
The expressions for the spectrum are complete to order $O(\epsilon^2)$ for all relevant couplings and, as we will show in section~\ref{multi-critical}, also for the marginal ones. 
For irrelevant couplings, due to mixing effects, only the $O(\epsilon)$ terms are complete and correctly agree with the CFT results~\cite{Codello:2017qek}.

From the analysis of the quadratic part of the beta function we find the following form for the universal OPE coefficients in the $\overline{\rm MS}$ scheme
\begin{equation}
 \begin{split}
  \tilde{C}^k{}_{ij}&
  = \frac{1}{2}i(i-1) j(j-1) \left(1-\frac{1}{6}(i j-4) \epsilon-\frac{17}{162} (i j-4)   \epsilon ^2\right) \delta _{4,i+j-k}
   \\&
  +\frac{8}{3} \epsilon \left(1   
   +17 \epsilon \right) \delta _{4,i} \delta _{4,j} \delta _{4,k}
  + \frac{2}{3} \epsilon \left( 1  +\frac{17}{27} \epsilon \right) (i \delta _{i,k}\delta _{4,j}
  +j \delta _{j,k} \delta_{4,i})\,.
  \label{Cijk-ising}
 \end{split}
\end{equation}
This general expression gives us back the results of our previous example \eqref{Cising} and thus matches, when overlapping, with CFT computations~\cite{Codello:2017qek},
but its general range of validity will become clearer in section~\ref{multi-critical} after we analyze the effects of mixing.
We note that a NNLO computation, beside bringing some mixing effects, 
will provide further contributions at order $O(\epsilon^2)$ so that one should consider at this level of accuracy 
the expressions \eqref{Cijk-ising} just \emph{up to order} $O(\epsilon)$ as in Eq.~\eqref{Cising}.
In fact, recalling our discussion in section~\ref{transformation} on the possible differences among coefficients computed in other schemes, 
agreement at order $O(\epsilon^2)$ with an NNLO computation can be observed only if the $\overline{\rm MS}$ scheme 
and the ``CFT'' scheme are related by a Hessian of order $O(\epsilon^2)$.

\subsection{{\tt Lee-Yang} universality class in LPA}\label{Lee-Yang}

The {\tt Lee-Yang} universality class has upper critical dimension $d_c=6$ and the LPA beta functionals at NLO for the dimensionful potential and wavefunction are
\begin{equation} \label{bvz_LY}
 \begin{split}
  \beta_V& = -\frac{1}{6}\frac{(V^{(2)})^3}{(4\pi)^3}-\frac{23}{144}\frac{(V^{(2)})^3(V^{(3)})^2}{(4\pi)^6}
 \\
  \beta_Z &= -\frac{1}{6}\frac{(V^{(3)})^2}{(4\pi)^3}-\frac{13}{216}\frac{(V^{(3)})^4}{(4\pi)^6}\,.
 \end{split}
\end{equation}
The explicit derivation of these beta functionals is quite straightforward.
After the convenient rescaling of the potential $v \!\to \!2(4\pi)^{3/2}\, v$ the beta functionals for dimensionless quantities are
\begin{equation}
 \begin{split}
  \beta_v &=
  -6 v + 2\varphi v^{(1)}
  +\epsilon \left(v -\frac{1}{2}\varphi v^{(1)}\!\right)
  +\frac{1}{2}\eta\varphi v^{(1)}
  -\frac{2}{3}(v^{(2)})^3-\frac{23}{9}(v^{(2)})^3(v^{(3)})^2
  \\
  \beta_z &=
  \eta z + 2\varphi z^{(1)}
  -\frac{\epsilon}{2}\varphi z^{(1)}
  +\frac{1}{2}\eta\varphi z^{(1)}
  -\frac{2}{3}(v^{(3)})^2-\frac{26}{27}(v^{(3)})^4\,.
  \label{LPALee}
   \end{split}
\end{equation}
Expanding the potential as Eq.~\eqref{eq:potential_couplings} 
leads to the general expression for the beta functions
\begin{equation}
 \begin{split}
  \beta_k =&
   -\!\left(6-2k-\Big(1-\frac{k}{2}\Big) \epsilon -3kg_3^2\left( 1+13 g_3^2 \right)\right)g_k
  \\&
  -\frac{2}{3} \sum_{i=2}^{k+4} \,\,\, \sum_{j=2}^{k+4-i} \, i(i-1) j(j-1) (i+j-k-6) (i+j-k-5) g_i \,g_j\, g_{6-i-j+k}
  \\&
  +\frac{23}{90} \sum_{i=2}^{k+10}\,\, \sum_{j=2}^{k+10-i}\,\, \sum_{t=2}^{k+10-i-j}\,\, \sum_{u=2}^{k+10-i-j-t}\, i(i-1) j(j-1) t(t-1) u(u-1) \times
  \\&
  \times (I-k-12) (I-k-11) \Big( 56 + 8 k - (k+12)  I + i j + i t + i u + j t + j u + t u + J  \Big) \times
  \\ & \times g_i \,g_j\, g_t\, g_u\,g_{k-I+12}
\end{split}
\end{equation}
where $I=i+j+t+u$ and $ J= i^2 + j^2 + t^2 + u^2$.
From \eqref{LPALee} we also immediately obtain the anomalous dimension
\begin{equation}
 \eta 
 =24 g_3^2 + 1248 g_3^4\,.
 \label{etaLee}
\end{equation}
As expected, because of the non-unitarity of the model, the fixed point is complex
\begin{equation}\label{LY-fp}
  g_k^* = g\,\delta_{k,3} \qquad\qquad
  g =\frac{1}{6\sqrt{6}} (-\epsilon)^{1/2}-\frac{125}{1944\sqrt{6}}(-\epsilon)^{3/2} + O\Big((-\epsilon)^{5/2}\Big)\\
    \,,
\end{equation}
showing that the $\epsilon$-expansion for the {\tt Lee-Yang} universality class is in fact an expansion in powers of $\sqrt{-\epsilon}$;
equivalently one can write $g^2 = -\frac{1}{54}\epsilon -\frac{125}{8748}\epsilon^2 +O(\epsilon^3)$.

After expanding the beta functions around the fixed point we determine the spectrum in terms of the critical coupling
\begin{equation}
 \begin{split}
  \theta_i
&  =
  6-2i-\left(1-\frac{i}{2}\right) \epsilon
  -\!\left(1-\frac{2g^2}{9}\right)\frac{7g^2}{12}i
   \\& 
  \qquad\qquad\qquad+\!\left(1-\frac{23}{24} g^2\right)\frac{g^2}{2} i^2
     +\! \frac{23}{72}g^4i^3
  -\!\left(\frac{g^2}{2}+\frac{13}{36}g^4\right)\delta_{i,3}\,,
\end{split}
\end{equation}
from which using \eqref{theta} we can extract the anomalous dimensions, as a function of $\epsilon$ and taking in account \eqref{LY-fp}
\be \label{gamma_LY}
\tilde \gamma_i = \frac{1}{18} i (6 i-7) \epsilon-\frac{1}{2916}i(414i^2-1371i+1043)\epsilon^2-\left(\frac{\epsilon}{3}+\frac{47}{486}\epsilon^2\right)\delta_{i,3}\,.
\ee
For reference we write the first anomalous dimensions
$$
\tilde\gamma_1 =-\frac{\epsilon}{18} -\frac{43}{1458}\epsilon^2\qquad\qquad \tilde \gamma_2 =\frac{5}{9}\epsilon +\frac{43}{1458}\epsilon^2 \qquad\qquad \tilde \gamma_3 =\frac{3}{2} \epsilon-\frac{125}{162}\epsilon^2\,.
$$
It is easy to check that the scaling relation $\theta_1+\theta_2 = d$, discussed in appendix~\ref{genscaling}, is indeed satisfied.

For the universal $\overline{\rm MS}$ OPE coefficients we obtain
\begin{equation}
 \begin{split}
  \tilde{C}^k{}_{ij}
  =& -12  i(i-1)j(j-1)g\left\{1 + 46  ( i + j + i j-5) g^2\right\}\delta_{i+j,k+3}
  \\&
  +36 g (1 + 312 g^2)\delta_{i,3}\delta_{j,3}\delta_{k,3}
  +12 g (1 + 104 g^2)\left(i\delta_{j,3}\delta_{i,k}+j\delta_{i,3}\delta_{j,k}\right)\,,
  \label{Cijk-LY}
 \end{split}
\end{equation}
which at this order we display as a function of the coupling $g$ of \eqref{LY-fp} for notational simplicity.
Using the explicit for of the fixed point \eqref{LY-fp} as a function of $\epsilon$,
and considering only the leading order in $\sqrt{-\epsilon}$ we find
\begin{equation}
 \begin{split}
  \tilde{C}^k{}_{ij}
  =\sqrt{\frac{2}{3}}\sqrt{-\epsilon} \,i(i-1)j(j-1)\delta_{i+j,k+3}
  +\sqrt{-6\epsilon}\,\delta_{i,3}\delta_{j,3}\delta_{k,3}
  +\sqrt{\frac{2}{3}}\sqrt{-\epsilon}\,\left(i\delta_{j,3}\delta_{i,k}+j\delta_{i,3}\delta_{j,k}\right)\,.
 \end{split}
\end{equation}
The first two universal OPE coefficients are
\begin{equation}
\tilde{C}^{1}{}_{2\,2} = - 4\sqrt{\frac{2}{3}}\sqrt{-\epsilon}
\qquad\qquad
\tilde{C}^{1}{}_{1\,3} = \sqrt{\frac{2}{3}}\sqrt{-\epsilon}
\end{equation}
and agree with CFT computations~\cite{Codello:2017qek}.
The discussion of the universality of the {\tt Ising}'s OPE coefficients has an analog here:
In the case of the {\tt Lee-Yang} universality class we have that $(\theta_c - \theta_a - \theta_b)=O(\sqrt{\epsilon})$,
therefore the eventual Hessian relating the $\overline{\rm MS}$ and CFT schemes
might contribute by changing the universal OPE coefficients at $O(\epsilon)$ or higher.
%

\section{Functional perturbative RG and the derivative expansion} \label{sec de}

In the rest of the paper we would like to show how it is possible to generalize the results presented so far to arbitrary order in the $\epsilon$-expansion
to include mixing effects, and also to extend the analysis to a wider set of universality classes. 
In order to enter into this subject and also pave the way for future computations including the most general operators, we will first describe the general setup of the {\it derivative expansion}
where one can systematically include higher-derivative operators. 
In what follows, our aim would be to outline a systematic approach to such a derivative expansion in the functional perturbative RG.
The derivative expansion, although being formally a truncation of the most general action \eqref{LPA}, 
allows, when combined with the perturbative $\epsilon$-expansion, a systematic and complete determination of the $\epsilon$-series
of the spectrum and the $\overline{\rm MS}$ OPE coefficients.

At each order in the number of derivatives there is an infinite number of operators with higher and higher powers of the field. 
Just like the potential function $V(\phi)$ which encompasses an infinite set of couplings, 
the couplings of these derivative operators can be collected into functions so that at each derivative order 
there is a finite basis of ``functional'' operators which spans all the operators with the given number of derivatives. 
To make it more explicit, one can denote the basis of functional operators with $k$ derivatives by $\hat W^{(k)}_a(\phi)$, where $a$ runs from $1$ to $N_k$, the number of elements in such a basis. With this notation the action \eqref{generalS} can be re-expressed  as
\be \label{de} 
S= \sum_{k\geq 0}\sum_{a=1}^{N_k} \int {\rm d}^d x \,\hat W^{(k)}_a\!(\phi)\,,
\ee
where the index $k$ runs over the number of derivatives and $a$ spans the possible degeneracy.

The first few instances of such operators can be listed as follows:
\\
\begin{equation}
\label{basis}
\hat W^{(0)}_1\!(\phi) = W^{(0)}_1\!(\phi) \qquad\qquad
\hat W^{(2)}_1\!(\phi) = W^{(2)}_1\!(\phi){\textstyle{\frac{1}{2}}} (\partial\phi)^2\nonumber
\end{equation}

\begin{equation}
\hat W^{(4)}_1\!(\phi) = W^{(4)}_1\!(\phi) (\Box\phi)^2  \qquad
\hat W^{(4)}_2\!(\phi) = W^{(4)}_2\!(\phi) \Box\phi(\partial\phi)^2 \qquad
\hat W^{(4)}_3\!(\phi) = W^{(4)}_3\!(\phi) (\partial\phi)^4 \,.\nonumber
\end{equation}
Thus $N_0 = 1$, $N_2 = 1$ and $N_4=3$.  Obviously, $V=W^{(0)}_1$, $Z=W^{(2)}_1$ and we will adopt the notation $W_a \equiv W^{(4)}_a$. 
One can continue in this way and choose a basis for higher-derivative operators. 
At the next order, i.e.\ six derivatives, there are $N_6=8$ independent functional operators which form a basis. This will increase to $N_8=23$ for the case of eight derivatives, and so on.
For each of the operators in \eqref{de}, after shifting to the relative dimensionless functionals $w^{(k)}_a$, 
one can define a (dimensionless) beta functional $\beta^{(k)}_a$ which captures the flow of the infinite number of couplings in $w^{(k)}_a$.  

At this stage, let us be more specific and concentrate on theories of the form \eqref{de}  close to the upper critical dimensions of multicritical $\phi^{2n}$ models \cite{Codello:2017qek}
\be 
d = \frac{2n}{n-1} - \epsilon\,.
\label{dncritica}
\ee
Then dimensional regularization has the virtue that the fixed-point action solving $\beta^{(k)}_a = 0$ is extremely simple,
since all fixed point functionals are zero apart from the potential $V$, which in turn is proportional to the critical coupling, and the $Z$, which is constant and can be set to one. 
The fixed point action is then
\be 
S_*= \int {\rm d}^d x \left\{ {\textstyle{\frac{1}{2}}}(\partial\phi)^2 + g\,\phi^{2n} \right\}\,.
\label{SFP}
\ee
These choices define the multi-critical universality classes which will be discussed in detail in the next section.

The action \eqref{de} can then be seen as a deformation around the fixed point \eqref{SFP} away from criticality.
One can formally define the stability matrix and the set of OPE coefficients in a functional form by expanding the beta functionals around the fixed point 
as\footnote{More generally, for an arbitrary Lagrangian $\mathcal{L}$, the RG flow can be formally described by a beta functional $\beta[\mathcal{L}]$, 
and a fixed point $\mathcal{L}_*$ of the theory would be defined by the condition $\beta[\mathcal{L}_*]=0$. 
The fixed point Lagrangian $\mathcal{L}_*$ is normally expected to describe a CFT, whenever scale invariance implies conformal invariance. 
Several non trivial informations on the critical theory can then be extracted by probing arbitrary off-critical deformations from the fixed point parametrized by $\mathcal{L}=\mathcal{L}_*+\delta \mathcal{L}$
\be \label{beta_functional}
\beta[\mathcal{L}_*+\delta \mathcal{L}] = \left.\frac{\delta \beta}{\delta \mathcal{L}}\right|_{\mathcal{L}_*} \delta \mathcal{L} +  \frac{1}{2}\left. \frac{\delta^2 \beta}{\delta \mathcal{L}\delta \mathcal{L}}\right|_{\mathcal{L}_*} \delta \mathcal{L} \delta \mathcal{L}+ \cdots
\ee
}
\be 
\beta^{(k)}_a( w_*+\delta w) = \sum_i\frac{\delta \beta^{(k)}_a}{\delta w^{(i)}_b} \Big|_*\delta w^{(i)}_b +  \frac{1}{2} \sum_{ij}\frac{\delta^2 \beta^{(k)}_a}{\delta w^{(i)}_b \delta w^{(j)}_c} \Big|_*  \delta w^{(i)}_b\delta w^{(j)}_c+ \cdots
\label{FRGbeta}
\ee
In the above expression of the beta for the functions $w^{k}_a$, which depend non linearly also on derivative of them, 
one has formally functional derivatives and integral are understood when repeated indices $a,b,\cdots$ are present.
Although one can study the RG flow and compute all universal quantities directly at the functional level by exploring the consequences of \eqref{FRGbeta},
in the next section we will reconnect with the discussion in terms of couplings as outlined in section~\ref{gen_RG_CFT}, 
and use the beta functionals $\beta^{(k)}_a$ as a convenient way to generate the coupling beta functions.
 
The couplings in \eqref{de} can be defined by expanding the functions such as $V(\phi)$, $Z(\phi)$, $W_a(\phi)$ and those of the higher derivative operators, 
in powers of the field, starting with $\phi^0=1$. 
Using dimensional analysis and recalling that close to the upper critical dimension the spectrum of the theory is almost Gaussian,
we can infer that the couplings in $V(\phi)$ corresponding to the $2n$ lowest dimensional operators $1,\phi,\cdots, \phi^{2n-1}$ do not mix with any other coupling.
Staring from $\phi^{2n}$ and all the way up to $\phi^{4n-3}$ they mix with the $O(\partial^2)$ couplings of $(\partial\phi)^2, \cdots, \phi^{2n-3}(\partial\phi)^2$. From $\phi^{4n-2}$, $\phi^{2n-2}(\partial\phi)^2$ the $O(\partial^4)$ couplings of $W_a(\phi)$ will also be involved. This can be summarized in the following table
\\
\be \label{mix}
\ba{rlllllllllll}
V: & \hspace{20pt}1 & \phi & \cdots & \phi^{2n-1} &  \phi^{2n} & \cdots & \phi^{4n-3} & \phi^{4n-2} & \phi^{4n-1} & \phi^{4n} & \cdots \\[2pt]
Z: &&&&& (\partial\phi)^2 & \cdots & \phi^{2n-3}(\partial\phi)^2 & \phi^{2n-2}(\partial\phi)^2 & \phi^{2n-1}(\partial\phi)^2 & \phi^{2n}(\partial\phi)^2 & \cdots \\[2pt]
W_1: &&&&&&&& (\Box\phi)^2 & \phi(\Box\phi)^2 & \phi^2(\Box\phi)^2 & \cdots \\[2pt]
W_2: &&&&&&&&& \Box\phi(\partial\phi)^2 & \phi\Box\phi(\partial\phi)^2  & \cdots \\[2pt]
W_3: &&&&&&&&&& (\partial\phi)^4 & \cdots 
\ea
\ee
\\
where each row collects the operators included in the function shown on the left-hand side and only couplings of operators in the same column mix together.
If we arrange the couplings of \eqref{de} in increasing order of their canonical operator dimension,
and furthermore, we sort them for increasing order of derivatives of their corresponding operators, the stability matrix takes the block-diagonal form
\be \label{sm}
\left(
\ba{cccc} 
M^{(0)} & & & \\  
& M^{(2)} &  & \\
& & M^{(4)} & \\ 
& & & \ddots
\ea
\right)
\ee
where in general $M^{(2k)}$ is itself a block diagonal matrix.
Each diagonal block contained in $M^{(2k)}$ describes the mixing between couplings of operators up to $2 k$ derivatives, 
all of which belong to the same column in \eqref{mix}.
In particular $M^{(0)}$ is a diagonal matrix with entries giving the scaling dimensions of the first $2n$ couplings in the potential.
The matrix $M^{(2)}$ is block diagonal, with each block being a two by two matrix which gives the mixing between a coupling in $V(\phi)$ and a coupling in $Z(\phi)$. 
$M^{(4)}$ is also a block diagonal matrix of which, with our choice of basis for the four-derivative operators, the first block is a three by three matrix, the second is four by four and the rest are five by five matrices.

Using dimensional analysis one can restrict the stability matrix even further if one is satisfied with the order $\epsilon$ approximation. 
As we will show explicitly in the following section, at this order each diagonal block in the matrices $M^{(2k)}$ in \eqref{sm} 
is itself block lower-triangular, where each block describes the mixing of operators with the same number of derivatives. This ensures that the entries on the diagonal
for the couplings of the potential and the second-derivative operators will give the scaling dimensions and are unaffected by the mixing at this order. 

\section{General $\phi^{2n}$ universality class} \label{multi-critical}

After the analysis of the {\tt Ising} universality class in sections~\ref{ising0} and \ref{ising1} and the introductory discussion of the previous sections,
we are now in a position to extend these results to general  models with even interaction $\phi^{2n}$ at the critical point.
In fact one can treat the whole set of universality classes $\phi^{2n}$ in a unified framework. A brief review of the method which closely follows \cite{ODwyer:2007brp}
is outlined in appendix~\ref{osborn}. Here we pick the main results that will be needed for our analysis. 

Throughout this work we will not go beyond second order in the derivative expansion, and in fact mostly concentrate on the local potential approximation. 
Let us therefore consider a theory of the form 
\be 
S = \int {\rm d}^d x \left\{ {\textstyle{\frac{1}{2}}} Z(\phi) (\partial\phi)^2 + V(\phi) \right\} \,,
\ee
in a space-time dimension which is close,  as in Eq.~\eqref{dncritica}, to the upper critical dimension at which the coupling of the interaction $\phi^{2n}$ becomes dimensionless.

The propagator of this theory satisfies the differential equation $-\square\, G_x = \delta^d_x$, where $\delta^d_x$ is the $d$-dimensional Dirac delta function. The solution is given by 
\be \label{prop}
G_x = \frac{1}{4\pi} \frac{\Gamma(\delta)}{\pi^\delta} \frac{1}{|x|^{2\delta}}\,, 
\ee
which is more conveniently written in terms of the field dimension $\delta = \frac{d}{2} -1 = \frac{1}{n-1} -\frac{\epsilon}{2} $. The coefficient in the propagator evaluated at criticality appears many times in the calculations. For convenience we therefore call it $c$ from now on  
\be 
c \equiv \frac{1}{4\pi} \frac{\Gamma(\delta_n)}{\pi^{\delta_n}} \qquad \qquad\qquad
\delta_n = \frac{1}{n-1}\,.
\ee
Let us neglect for the moment the effect of derivative interactions encoded in $Z(\phi)$ and concentrate on the $V(\phi)$ contributions to the beta functions of $V(\phi)$ and $Z(\phi)$. Before giving the explicit expressions for the beta functions let us mention that in this $O(\partial^0)$ truncation one can extract the scaling dimensions and $\overline{\rm MS}$ OPE coefficients for the relevant components as they will be in any case unaffected by the mixing with the derivative operators. Moreover, as we will argue later, remaining within the same truncation it is possible to go beyond the relevant components if one is content with the order $\epsilon$ estimates.

Neglecting derivative interactions, the beta functional of the dimensionless potential, in the form of Eq.~\eqref{betav}, at the NLO (cubic order) in the dimensionless potential is 
{\setlength\arraycolsep{2pt}
\bea \label{bv}
\beta_v &=& -\,d\, v(\varphi) + \frac{d-2+\eta}{2} \,\varphi \, v'(\varphi)+ \frac{n-1}{n!}\frac{c^{n-1}}{4} v^{(n)}(\varphi)^2 \nn\\[12pt]
&& -\frac{n-1}{48}\,c^{2n-2}\,\Gamma(\delta_n)\hspace{-0.03\textwidth}\sum_{{\footnotesize \ba{c}r\!+\!s\!+\!t\!=\!2n\\[-5pt] r,s,t \neq n\ea}}\hspace{-0.02\textwidth}\frac{K^n_{rst}}{r!s!t!} \;v^{(r+s)}(\varphi)\,v^{(s+t)}(\varphi)\,v^{(t+r)}(\varphi) \nn\\
&& -\frac{(n-1)^2}{16 \,n!}\,c^{2n-2}\,\hspace{-0.01\textwidth}\sum_{\footnotesize s+t=n}\frac{n-1+L^n_{st}}{s!t!} \;v^{(n)}(\varphi)\,v^{(n+s)}(\varphi)\,v^{(n+t)}(\varphi)\,,
\eea}%
where the integers $r,s,t$ are implicitly taken to be positive, and the quantities $K^n_{rst}$ and $L^n_{st}$ are defined as follows
\be 
K^n_{rst} = \frac{\Gamma\left(\frac{n-r}{n-1}\right)\Gamma\left(\frac{n-s}{n-1}\right)\Gamma\left(\frac{n-t}{n-1}\right)}{\Gamma\left(\frac{r}{n-1}\right)\Gamma\left(\frac{s}{n-1}\right)\Gamma\left(\frac{t}{n-1}\right)}\,, \qquad 
L^n_{st} = \psi(\delta_n) - \psi(s\delta_n)-\psi(t\delta_n) + \psi(1)\,, 
\ee
where $\psi(x)=\Gamma'(x)/\Gamma(x)$ is the digamma function.
The last term in the first line of \eqref{bv} is the LO $(n-1)$-loop term, while the NLO second and third lines appear at $2(n-1)$-loops. The origin of such terms and the corresponding diagrams will be briefly discussed in appendix~\ref{osborn}. 
Notice also that, differently from sections \ref{gen_RG_CFT} and \ref{section-tutorials}, we did not yet include any further rescaling when moving from the dimensionful $V(\phi)$ to the dimensionless $v(\varphi)$ potential:
since the rescaling does not affect the spectrum, we postpone the discussion of the ``appropriate'' rescaling to subsection \ref{OPE} in which some $\overline{\rm MS}$ OPE coefficients are computed.

Neglecting derivative interactions (in agreement with our definition of LPA), the induced flow of the function $z(\varphi)$ at quadratic order
is given by
\be \label{bz}
\beta_z = \eta\, z(\varphi) + \frac{d-2+\eta}{2}\, \varphi \, z'(\varphi) -  \frac{(n-1)^2}{(2n)!}\frac{c^{2n-2}}{4}\, v^{(2n)}(\varphi)^2.
\ee
The last term in this equation comes from a diagram with $2(n-1)$-loops, which gives a counter-term consisting of the second contribution in Eq.~\eqref{lct}, as explained in appendix~\ref{osborn}. 

From \eqref{bv}, noticing the fact that only the dimensionless coupling can take a non-zero value at the fixed point, one can set $v(\varphi) = g\,\varphi^{2n}$ together with the condition $\beta_v =0$ to find the critical coupling $g$ at quadratic order in $\epsilon$. This is given by
\be  \label{fp}
\frac{(2n)!^2}{4\,n!^3} \, c^{n-1} g = \epsilon - \frac{n}{n\!-\!1}\eta  + \frac{n!^4}{(2n)!}\bigg[\frac{1}{3}\,\Gamma(\delta_n)\;n!^2\hspace{-0.04\textwidth}\sum_{{\footnotesize \ba{c}r\!+\!s\!+\!t\!=\!2n\\[-5pt] r,s,t \neq n\ea}}\hspace{-0.02\textwidth}\frac{K^n_{rst}}{(r!s!t!)^2}  +(n\!-\!1)\,\hspace{-0.015\textwidth}\sum_{\footnotesize s+t=n}\!\!\frac{n\!-\!1+L^n_{st}}{s!^2t!^2} \bigg]\epsilon^2\,.
\ee
Notice that here we have used an expansion of $v(\varphi)$ without factorials. Including the factorials, the $(2n)!$ on the left-hand side would have appeared with power one, in agreement with \cite{ODwyer:2007brp,Codello:2017qek}. This, of course, does not affect the final physical results when written in terms of $\epsilon$.

The anomalous dimension can be read off from \eqref{bz} imposing the condition $\left.\beta_z\right|_{\varphi=0}=0$ and using $z(0)=1$. This gives, after using \eqref{fp},
\be 
\eta = \frac{4(n-1)^2 n!^6}{(2n)!^3}\epsilon^2\,,
\ee
in agreement with \cite{ODwyer:2007brp,Itzykson:1989sx} and recent CFT based computations~\cite{Codello:2017qek, Gliozzi:2016ysv}.
Having at our disposal the functional form of $\beta_v$ at cubic order, we can follow the prescription of section~\ref{gen_RG_CFT} to find the scaling dimensions of the relevant couplings at $O(\epsilon^2)$ and the $\overline{\rm MS}$ OPE coefficients for the relevant operators at $O(\epsilon)$. 

However, before doing so let us devote the next subsection to considering the leading order mixing effects due to the presence of $z(\varphi)$-interactions. This, for instance, will allow us to compute the leading order anomalous scaling dimensions of the $z(\varphi)$ couplings, and justify the validity of the leading order anomalous dimensions of the $v(\varphi)$ couplings.


\subsection{Mixing}

One can take into account the mixing effects due to the presence of derivative operators. We have given a general sketch of the mixing pattern in section~\ref{sec de}. 
Here we concentrate on explicit results for two-derivative interactions collected in the function $z(\varphi)$. 
At quadratic level the presence of $z(\varphi)$ does not affect the beta function of the potential \eqref{bv}, but \eqref{bz} instead gets a contribution at this level 
\be  \label{bz2}
\Delta\beta_z =  \frac{n-1}{n!}\frac{c^{n-1}}{2} \left[z^{(n)}(\varphi)\,v^{(n)}(\varphi)+z^{(n-1)}(\varphi)\,v^{(n+1)}(\varphi)\right]\,. 
\ee
These new terms arise from the derivative interactions and appear at $(n-1)$-loops~\cite{ODwyer:2007brp} as discussed in appendix~\ref{osborn}. 
This gives rise to a mixing, at order $\epsilon$, between the operators $\phi^{k+2n}$ and $\phi^k(\partial\phi)^2$ for $k=0,\cdots,2n-3$, 
described by the $k$-th block of the matrix $M^{(2)}$, while the couplings of $\phi^k$ with $k=0,\cdots,2n-1$, and therefore the elements of $M^{(0)}$, are unaffected.

In general the terms in a beta function which contribute to the stability matrix at order $\epsilon$ must be quadratic in the couplings, 
and furthermore one of the couplings must be the dimensionless coupling $g$ which is the only one that takes a nonzero value at the fixed point. 
In the beta functional this manifests as the product of the potential $v(\varphi)$ and a function corresponding to a higher derivative operator, 
or more precisely, the product of a derivative of these functions. 
These terms come from diagrams of the form displayed in  Fig.~\ref{zmelon} of appendix \ref{osborn}, or its generalizations where instead of $z(\varphi)$ one can have functions encoding higher derivative interactions. 

A simple argument based on dimensional analysis shows that generally in the beta function of a $2k$-derivative coupling, 
the quadratic term which includes the coupling $g$ can involve also a derivative coupling lower or equal to $k$. 
To show this, one should notice that in dimensional regularization
the diagrams contributing to the beta functions must be dimensionless, 
i.e.\ have vanishing superficial degree of divergence. 
For a melon diagram of the form in Fig.~\ref{zmelon} with $r$ propagators that includes the potential $v(\varphi)$ at one vertex 
and a $2l$-derivative coupling on the other vertex this condition is

\be 
(r-1)\frac{2n}{n-1}-2r+2(k-l) =0\,.
\ee
The first term comes from the $r-1$ loop integrations, while the $r$ propagators give a contribution $-2r$ in the second term. To justify the remaining terms one should notice that there are altogether $2l$ derivatives at one of the vertices, some of which might act on the propagators and some might not, but finally we would like to extract the $\partial^{2k}$ contribution from this diagram. This leads to the contribution $2(k-l)$. This simple relation can be re-arranged and put in a more useful form,
\be  \label{0}
l-k = \frac{n-r}{n-1} <1\,,
\ee
where one can use the fact that $r\ge 2$ on the right-hand side to put an upper bound on $l-k$. Since $k,l$ are integers, the inequality \eqref{0} says that $l\leq k$, which is the statement claimed above. This is a more general case of what we have already seen: that the beta functions of the potential couplings do not contain the product of $v$ and $z$ couplings but only $v$-coupling squared. This is telling us that at order $\epsilon$ each diagonal block in the stability matrix that describes the mixing of a column in \eqref{mix} is itself block lower-triangular (where here a block describes the mixing of operators with the same number of derivatives), as will be shown explicitly in the simplest case in the next section. This ensures that at order $\epsilon$ the eigenvalues of the potential and the two-derivative couplings are never affected by the mixing. In particular from \eqref{bv} one can find the spectrum of the couplings in the potential, not only for the relevant ones but also for the marginal and all the irrelevant couplings. Similarly, the beta function \eqref{bz} with the correction \eqref{bz2} gives the spectrum of all the $z$-couplings, at order $\epsilon$. These are made more explicit in the following subsection.

\subsection{Spectrum}\label{spectrum-subsection}

In order to proceed with explicit results let us stick to the following convention throughout  this section for the expansions of the functions $v(\varphi)$, $z(\varphi)$ and of the corresponding beta functionals in powers of the field
{\setlength\arraycolsep{2pt}
\be  \label{expansions}
\ba{l}
v(\varphi) =\displaystyle \sum_{k=0} g_k \varphi^k \\[20pt]
z(\varphi) = \displaystyle \sum_{k=0} h_k \varphi^k
\ea \qquad
\ba{lll}
\beta_v(\varphi) = \displaystyle \sum_{k=0} \beta_v^k \varphi^k \\[20pt]
\beta_z(\varphi) = \displaystyle \sum_{k=0} \beta_z^k \varphi^k.
\ea
\ee}%
The choice of normalization for the couplings is of course physically irrelevant.
An explicit computation using the beta function \eqref{bv} and the expansions \eqref{expansions} shows that the matrix $\partial\beta^i_v/\partial g_j$ evaluated at the fixed point, which is for dimensional reasons diagonal, has the elements $-\theta_i= -d + i(d-2)/2+\tilde\gamma_i$ on its diagonal, with the following anomalous 
parts\footnote{Note that in order to be able to make sense of the formula for the anomalous dimensions $\tilde\gamma_i$ for general $i$, 
the terms involving factorials of negative numbers in the denominators are interpreted to be zero by analytic continuation.}
{\setlength\arraycolsep{3pt}
\bea \label{spectrum}
\tilde\gamma_i &=& i\frac{\eta}{2} 
+ \frac{(n-1)i!}{(i-n)!}\,\frac{2\,n!}{(2n)!}\left[\epsilon - \frac{n}{n-1} \,\eta\right] + 2n\, \eta\, \delta^{2n}_i 
\nn\\[7pt]
&+& \frac{(n-1)i!n!^6}{(2n)!^2}\,\Gamma(\delta_n)\hspace{-0.03\textwidth}\sum_{{\footnotesize \ba{c}r\!+\!s\!+\!t\!=\!2n\\[-5pt] r,s,t \neq n\ea}}\hspace{-0.02\textwidth}\frac{K^n_{rst}}{(r!s!t!)^2} \left[\frac{2n!}{3(i-n)!} -\frac{r!}{(i-2n+r)!}\right]\epsilon^2 \nn\\[-3pt]
&+& \frac{(n-1)^2i! n!^5}{(2n)!^2}\,\hspace{-0.01\textwidth}\sum_{\footnotesize s+t=n}\frac{n-1+L^n_{st}}{(s!t!)^2}\left[\frac{1}{(i-n)!} -\frac{2s!}{n!(i-2n+s)!}\right]\epsilon^2.
\eea}%
For the relevant components, that is for the range $0\leq i \leq 2n-1$, these are simply the anomalous dimensions with accuracy $O(\epsilon^2)$. 
The last term in the first line, which comes from the term proportional to $\eta$ in \eqref{bv}, does not contribute in the relevant sector. 
However, if one wishes to find the anomalous dimension of the marginal coupling, one has to take this term into account.
Within the same $O(\epsilon^2)$ accuracy, for the irrelevant couplings, which we do not consider here, additional mixing transformations are required to diagonalize the stability matrix. 

From \eqref{spectrum} one can readily see that for $i=1$ all the terms except the first vanish.
Also, interestingly, for $i=2n-1$ which corresponds to the descendant operator $\phi^{2n-1}$ in the interacting theory because of the Schwinger-Dyson equations, 
the $O(\epsilon^2)$ terms in the second and third line of Eq.~\eqref{spectrum} vanish so that these anomalous dimensions take the simple form 
\be 
\tilde\gamma_1= \frac{\eta}{2}, \qquad
\tilde\gamma_{2n-1} = (n-1)\epsilon -\frac{\eta}{2}\,.
\label{multianom}
\ee
The two anomalous dimensions then sum up to $\tilde\gamma_1 + \tilde\gamma_{2n-1} = (n-1)\epsilon$, which is equivalent to the scaling relation $\theta_1+\theta_{2n-1}=d$ and proved in general in appendix~\ref{genscaling}.

The correction \eqref{bz2} allows us to go beyond the local potential approximation and compute at order $\epsilon$ the block $M^{(2)}$ in \eqref{sm} which is a block-diagonal matrix with two by two blocks. The $i$-th block which gives the mixing of the $\phi^{i+2n}$ and $\phi^i(\partial\phi)^2$ couplings is given in the $\{ \phi^{i+2n},\phi^i(\partial\phi)^2\}$ basis as 
\be \label{mixing}
\frac{i}{n-1}\mathbf{1} + \left(
\ba{cc}
-\frac{(i+2n)}{2} +   \frac{2(n-1)n!}{(2n)!} \frac{(i+2n)!}{(i+n)!} & 0 \\[10pt]
- \frac{2(n-1)^2 n!^3}{(2n)!^2}\frac{(i+2n)!}{i!} c^{n-1} (1-\delta^i_0) &  -\frac{i}{2} +  \frac{2(n-1)n!}{(2n)!}\, \frac{(i+1)!}{(i-n+1)!}
\ea
\right)\epsilon \,+\, O(\epsilon^2) \,,
\ee
where $\mathbf{1}$ is the two dimensional identity matrix. 
For each $i$ the two eigenoperators have the same canonical scaling at the critical dimension. The eigenvalues of the stability matrix include the scaling dimensions $-\theta_{i+2n}$, given in Eq.~\eqref{theta}, and $(\frac{d}{2}-1)i+\tilde{\omega}_i$, which is the analog for $z$-couplings in the notation of \cite{ODwyer:2007brp}. 

From these, one can then read off the anomalous parts $\tilde{\gamma}_i$ and $\tilde{\omega}_i$ of the $v$ and $z$ coupling scaling dimensions at order $\epsilon$ which are valid not only for $0\leq i \leq 2n-3$ described by the above matrix but for all $i$, according to the discussion in the previous subsection.
In summary, again interpreting the factorials to be infinite for negative integer arguments, and for $i\geq 0$
\be
\tilde\gamma_i = \frac{2(n-1)n!}{(2n)!} \frac{i!}{(i-n)!} \, \epsilon  \qquad\qquad
\tilde\omega_i = \frac{2(n-1)n!}{(2n)!}\, \frac{(i+1)!}{(i-n+1)!} \, \epsilon  \,.\label{gammai}
\ee
%
This reproduces the result of \cite{ODwyer:2007brp}.
The $\tilde\gamma_i$ in Eq.~\eqref{gammai} also match the anomalous dimensions found in \cite{Codello:2017qek,Gliozzi:2016ysv} from CFT constraints.

Beyond the leading order for the anomalous dimensions, the stability matrix will not be lower-triangular anymore, and in order to find the anomalous dimensions of higher and higher powers of $\phi$ one has to take into account (up to cubic order contributions of) operators of higher and higher dimensions. In the simplest case, $2n<i<4n-3$, one needs to include cubic corrections to $\beta_z$, and furthermore, take into account the $z(\varphi)$ contribution to $\beta_v$ at cubic level. The only term contributing to this last piece is proportional to $v^{(n)}(\phi)^2 \ z(\phi)$ and leads to $O(\epsilon^2)$ corrections in the upper right element in \eqref{mixing}. These higher order corrections are not considered here and are left for future work. 

Besides \eqref{gammai}, an extra information which has been obtained in \cite{Codello:2017qek} using conformal symmetry and the Schwinger-Dyson equations is the leading order value of $\gamma_2$ for $n>2$, which is of order $\epsilon^2$. For $n>2$, putting $i=2$ in \eqref{spectrum} gives 
\be 
\tilde\gamma_2 = \eta -\frac{2(n-1)n!^6}{(2n)!^2}\,\Gamma(\delta_n)\,\frac{K^n_{2n-2,1,1}}{(2n-2)!}\,\epsilon^2  = \frac{8(n+1)(n-1)^3n!^6}{(n-2)(2n)!^3}\,\epsilon^2\,,
\ee
which is also in agreement with the result found in \cite{Codello:2017qek}.

\subsection{OPE coefficients}\label{OPE}

The only non-zero $\tilde{C}^k{}_{ij}$ coefficients that are extracted from the beta functions are those that are massless, or equivalently, satisfy the universality condition $i+j-k=2n$. Contrary to the anomalous dimensions, the OPE coefficients do depend on the normalization of the couplings.
Throughout this section we continue to use the normalization where couplings appear without factorials in the $v(\varphi),\, z(\varphi)$ expansions, as defined in \eqref{expansions}. On top of this, it turns out convenient to make a global rescaling of the couplings by redefining the potential according to\footnote{In Sections~\ref{ising1} and~\ref{Lee-Yang} this rescaling was used for both the {\tt Ising} and the {\tt Lee-Yang} universality classes with $n=1$ and $n=\frac{3}{2}$ respectively.}
\be 
v \rightarrow \frac{4}{(n-1)c^{n-1}}\,v\,.
\ee
This removes the parameter $c$ from the beta functions \eqref{bv} and \eqref{bz}. In such a normalization, using the beta function \eqref{bv}, the expansion of the potential and its beta functional in powers of the field \eqref{expansions}, and the fixed point relation \eqref{fp}, a lengthy but straightforward calculation based on the definition \eqref{n} gives the $\overline{\rm MS}$ OPE coefficients ($k=i+j-2n$)
{\setlength\arraycolsep{2pt}
\bea \label{ope}
\tilde{C}^k_{\;\,ij} &=& \frac{1}{n!}  \frac{i!}{(i-n)!}\frac{j!}{(j-n)!} -\Gamma(\delta_n) \frac{n!^3}{(2n)!}\hspace{-0.02\textwidth}\sum_{{\footnotesize \ba{c}r\!+\!s\!+\!t\!=\!2n\\[-5pt] r,s,t \neq n\ea}}\hspace{-0.02\textwidth}\frac{K^n_{rst}}{r!s!t!^2} \,\frac{j!}{(j-s-t)!}\frac{i!}{(i+s-2n)!}\,\epsilon \nn\\
&-& \frac{(n-1)n!^2}{(2n)!}\,\hspace{-0.01\textwidth}\sum_{\footnotesize s+t=n}\frac{n-1+L^n_{st}}{s!t!} \;\left[\frac{1}{n!}\frac{j!}{(j-n-s)!}\frac{i!}{(i-n-t)!}+\frac{1}{s!}\frac{i!}{(i-n)!}\frac{j!}{(j-n-s)!} \right. \nn\\[10pt]
&+& \left. \frac{1}{s!}\frac{j!}{(j-n)!}\frac{i!}{(i-n-s)!}\right]\epsilon + \frac{2(n-1)n!^3}{(2n)!}(i\,\delta^{2n}_j+j\,\delta^{2n}_i+2n\, \delta^{2n}_i\delta^{2n}_j)\epsilon\,.
\eea}%
Notice that, strictly speaking, the above quantity is in fact the matrix $N^k{}_{ij}$ defined in \eqref{n},
but because the mixing matrix ${\cal S}^i{}_a$ is diagonal (in the relevant and marginally irrelevant part of the spectrum) it coincides with the OPE coefficients $\tilde{C}^k{}_{ij}$ in our scheme.  
The last contribution in \eqref{ope} comes from the anomalous dimension term in \eqref{bv}. Similar to the anomalous dimensions \eqref{spectrum}
one has to keep in mind that terms with negative factorials in the denominators vanish. Notice that the first term is nothing but the combinatorial factor that comes from Wick contractions in the free theory. The normalization we have adopted therefore coincides with the CFT normalization where the coefficient of the two point function $\langle\phi\,\phi\rangle$ is set to unity. 

It is important to comment on the range of validity for the $i,j$ indices in the above formula. As in the case of anomalous dimensions, Eq.~\eqref{ope} is, of course, valid for all relevant components, that is, positive integer indices smaller than $2n$. Notice that in this case the last term does not contribute. However, this is not all we can extract from this formula. For instance, let us consider the case $i<n$.
The above formula will then be of order $\epsilon$. At this level of approximation Eq.~\eqref{ope} is valid for any $j$, and not only the relevant ones. This is because mixing effects enter only at NLO. 
Therefore for such cases one can use \eqref{bv} without any concern about the mixing. Notice that for these cases only the second term on the first line and the last term in Eq.~\eqref{ope} contribute. The particular case $\tilde{C}^1{}_{1,2n}$ gets contribution only from the last term in \eqref{ope} and takes the simple form  
\be \label{opemargin}
\tilde{C}^1{}_{1,2n} = \frac{2(n-1)n!^3}{(2n)!}\epsilon\,.
\ee
This reproduces the result found in \cite{Codello:2017qek} from CFT considerations, and therefore the computation done in the $\overline{\rm MS}$ scheme reproduces an entire family of CFT OPE coefficients at least at order $O(\epsilon)$.

Finally, let us consider Eq.~\eqref{ope} for $k=1$. The indices $i,j$ must then satisfy $i+j=2n-1$, so we choose them as $i=n-m$, $j=n+m+1$, for $m=1,\dots, n-1$. The OPE coefficients reduce to 
\be 
\tilde{C}^1{}_{n-m, n+m+1} =  \frac{(n-1)^2}{m(m+1)}\,\frac{(n+1+m)!(n-m)!}{(n-1-m)!(n+m)!} \,\frac{n!^3}{(2n)!}\epsilon\,.
\ee
This $\overline{\rm MS}$ result is also in agreement with \cite{Codello:2017qek}, and with \cite{Gliozzi:2016ysv} if one takes into account the different normalizations of the operators $\phi^l$.

\subsection{Examples of CFT data for specific theories}\label{examples}

Despite the above general treatment being comprehensive of all the even multicritical models,
we believe it is interesting to show some explicit results for specific theories. 
The case of {\tt Ising} had already been studied in subsections~\ref{ising0} and~\ref{ising1}. In this subsection we collect the CFT data for the {\tt Tricritical} and {\tt Tetracritical} universality classes. 

The {\tt Tricritical} universality class corresponds to $n=3$. The anomalous dimensions for the relevant and marginal operators at $O(\epsilon^2)$ can be obtained from the general formula \eqref{spectrum} and are given explicitly as 
\be 
\ba{rcl}
\tilde{\gamma}_1 &=& \displaystyle \frac{\epsilon^2}{1000} \\[9pt]
\tilde{\gamma}_2 &=& \displaystyle \frac{4\epsilon^2}{125} 
\ea \qquad
\ba{rcl}
\tilde{\gamma}_3 &=& \displaystyle \frac{\epsilon}{5} + \left(\frac{2037}{5000}+\frac{27\pi^2}{400}\right)\epsilon ^2 \\[5pt]
\tilde{\gamma}_4 &=& \displaystyle \frac{4\epsilon}{5} + \left(\frac{601}{625}+\frac{27\pi^2}{200}\right)\epsilon ^2
\ea \qquad
\ba{rcl}
\tilde{\gamma}_5 &=& \displaystyle 2\epsilon - \frac{\epsilon^2}{1000} \\[5pt]
\tilde{\gamma}_6 &=& \displaystyle 4\epsilon - \left(\frac{1689}{250}+\frac{27\pi^2}{40}\right)\epsilon^2\,.
\ea
\ee
The scaling relation $\tilde{\gamma}_1+\tilde{\gamma}_5=2\epsilon$ is satisfied. Notice that restricting to order $\epsilon$ one can immediately extend these results to all the couplings including the irrelevant ones, and even further to the $z(\varphi)$ couplings. These were reported in Eqs.~\eqref{gammai}. It is also easy to extract from the general equation \eqref{ope} the universal OPE coefficients in the $\overline{\rm MS}$ scheme with relevant components at $O(\epsilon)$. These are listed below
\be 
\ba{rcl}
\tilde C^1{}_{25} &=& \displaystyle 6\epsilon \\
\tilde C^1{}_{34} &=& \displaystyle 24-\frac{72}{5}\epsilon \\[7pt]
\tilde C^2{}_{35} &=& \displaystyle 60-90\epsilon
\ea \qquad
\ba{rcl}
\tilde C^2{}_{44} &=& \displaystyle 96-\frac{18}{5}(32+3\pi^2)\epsilon \\[6pt]
\tilde C^3{}_{45} &=& \displaystyle 240-6(98+9\pi^2)\epsilon \\[6pt]
\tilde C^4{}_{55} &=& \displaystyle 600-15(167+18\pi^2)\epsilon\,.
\ea
\ee
Furthermore Eq.~\eqref{ope} gives also the following leading order OPE coefficients with a marginal component
\be 
\tilde C^1{}_{16} = \frac{6}{5}\,\epsilon \qquad
\tilde C^2{}_{26} = \frac{192}{5}\,\epsilon\,,
\ee
and an infinite set of OPE coefficients with an irrelevant component
\be 
\tilde C^3{}_{27} = 126\,\epsilon \qquad
\tilde C^4{}_{28} = 336\,\epsilon \qquad
\tilde C^5{}_{29} = 756\,\epsilon \qquad \cdots
\ee
The OPE coefficients $\tilde C^1{}_{25}$ and  $\tilde C^1{}_{16}$ exactly match the corresponding structure constants computed with CFT methods in \cite{Codello:2017qek}. For the others there are no available CFT results to compare with.

For the {\tt Tetracritical} universality class, which corresponds to $n=4$, there are seven relevant couplings whose anomalous dimensions are
\be 
\ba{rcl}
\tilde{\gamma}_1 &=& \displaystyle \frac{9\epsilon^2}{171500} \\[8pt]
\tilde{\gamma}_2 &=& \displaystyle \frac{27\epsilon^2}{171500} \\[8pt]
\tilde{\gamma}_3 &=& \displaystyle \frac{7587\epsilon^2}{171500} 
\ea \qquad
\ba{rcl}
\tilde{\gamma}_4 &=& \displaystyle \frac{3\epsilon}{35} + \frac{3(477948+78400\,\Gamma[{\textstyle{\frac{1}{3}}}]^3+99225\log 3-33075\sqrt{3}\pi)\epsilon^2}{3001250} \\[5pt]
\tilde{\gamma}_5 &=& \displaystyle \frac{3\epsilon}{7} + \frac{9(232287+39200\,\Gamma[{\textstyle{\frac{1}{3}}}]^3+66150\log 3-22050\sqrt{3}\pi)\epsilon^2}{1200500} \\[5pt]
\tilde{\gamma}_6 &=& \displaystyle \frac{9\epsilon}{7} + \frac{3(646533+98000\,\Gamma[{\textstyle{\frac{1}{3}}}]^3+198450\log 3-66150\sqrt{3}\pi)\epsilon^2}{600250} \nn
\ea
\ee
\be 
\tilde{\gamma}_7 = \displaystyle 3\epsilon - \frac{9\epsilon^2}{171500}\,, 
\ee
while the anomalous dimension of the marginal coupling is given as
\be 
\tilde{\gamma}_8 = \displaystyle 6\epsilon - \frac{3(342516+39200\,\Gamma[{\textstyle{\frac{1}{3}}}]^3+99225\log 3-33075\sqrt{3}\pi)\epsilon^2}{42875}\,. 
\ee
As expected, the spectrum satisfies the scaling relation $\tilde{\gamma}_1+\tilde{\gamma}_7=3\epsilon$.
Using Eq.~\eqref{ope} we also list here, at order $\epsilon$, all the OPE coefficients with relevant components
\be 
\ba{rcl}
\tilde C^1{}_{27} &=& \displaystyle \frac{36}{5}\epsilon \\[9pt]
\tilde C^1{}_{36} &=& \displaystyle \frac{972}{35}\epsilon \\[5pt]
\tilde C^1{}_{45} &=& \displaystyle 120-\frac{432}{7}\epsilon 
\ea \qquad
\ba{rcl}
\tilde C^2{}_{37} &=& \displaystyle \frac{1566}{5}\epsilon \\
\tilde C^2{}_{46} &=& \displaystyle 360-\frac{324}{35}(5\sqrt{3}\pi-58-15\log 3)\epsilon \\[6pt]
\tilde C^2{}_{55} &=& \displaystyle 600-\frac{240}{7}(18+\Gamma[{\textstyle{\frac{1}{3}}}]^3)\epsilon
\ea
\ee
\be 
\ba{rcl}
\tilde C^3{}_{47} &=& \displaystyle 840-\frac{324}{5}\left(5\sqrt{3}\pi-38-15\log 3\right)\epsilon \\[8pt]
\tilde C^3{}_{56} &=& \displaystyle 1800-\frac{4}{7}\left(320\sqrt{3}\pi\Gamma[{\textstyle{\frac{1}{3}}}]^2\Gamma[{\textstyle{\frac{2}{3}}}]^{-1}+81(84+15\log 3-5\sqrt{3}\pi)\right)\epsilon \\[10pt]
\tilde C^4{}_{57} &=& \displaystyle 4200-12\left(1506+100\,\Gamma[{\textstyle{\frac{1}{3}}}]^3+405\log 3-135\sqrt{3}\pi\right)\epsilon \\[9pt]
\tilde C^4{}_{66} &=& \displaystyle 5400-\frac{2}{7}\left(4800\sqrt{3}\pi\Gamma[{\textstyle{\frac{1}{3}}}]^2\Gamma[{\textstyle{\frac{2}{3}}}]^{-1}+81(822+195\log 3-65\sqrt{3}\pi)\right)\epsilon \\[10pt]
\tilde C^5{}_{67} &=& \displaystyle 12600-\frac{54}{5}\left(7292+800\,\Gamma[{\textstyle{\frac{1}{3}}}]^3+2025\log 3-675\sqrt{3}\pi\right)\epsilon \\[10pt]
\tilde C^6{}_{77} &=& \displaystyle 29400-\frac{126}{5}\left(11519+1400\,\Gamma[{\textstyle{\frac{1}{3}}}]^3+3375\log 3-1125\sqrt{3}\pi\right)\epsilon\,,
\ea 
\ee
as well as some with a marginal component 
\be 
\tilde C^1{}_{18} = \frac{72}{35}\,\epsilon \qquad
\tilde C^2{}_{28} = \frac{432}{7}\,\epsilon \qquad
\tilde C^3{}_{38} = \frac{60696}{35}\,\epsilon\,,
\ee
the first of which can also be found from \eqref{opemargin}. We finally report here the first few of the infinite set of leading order OPE coefficients with an irrelevant component
\be 
\ba{llllllllll}
\tilde C^3{}_{2,9} &=& \displaystyle \frac{1296}{5}\,\epsilon & \qquad
\tilde C^4{}_{2,10} &=& \displaystyle 864\,\epsilon & \qquad
\tilde C^5{}_{2,11} &=& \displaystyle 2376\,\epsilon & \qquad \cdots \\[10pt]
\tilde C^4{}_{3,9} &=& \displaystyle \frac{33048}{5}\,\epsilon & \qquad
\tilde C^5{}_{3,10} &=& \displaystyle 20088\,\epsilon\, & \qquad
\tilde C^6{}_{3,11} &=& \displaystyle \frac{260172}{5}\,\epsilon & \qquad \cdots
\ea
\ee
For this universality class the OPE coefficients $\tilde C^1{}_{27} $, $\tilde C^1{}_{36}$ and $\tilde C^1{}_{18}$ correctly match the corresponding structure constants computed in \cite{Codello:2017qek}.
The considerations made for the {\tt Tricritical} case on the full comparison to CFT are equally valid for the {\tt Tetracritical} universality class.

\section{Conclusions} \label{sect-conclusions}

In this paper we have shown how to extend renormalization group (RG) techniques to the computation 
of some OPE coefficients at a scale invariant critical points of scalar quantum field theories. 
The approach of this work employs dimensional regularization in the $\overline{\rm MS}$ scheme
at the functional level and gives access to a specific set of ``massless'' OPE coefficients,
which are related to terms in the beta functions that are universal at the upper critical dimensions of the models under investigation. 
For general multicritical models we have extracted these quantities,
and we have shown that at order $O(\epsilon)$ they agree with the corresponding OPE coefficients computed directly
with CFT methods, when available on both sides.

Let us briefly summarize our procedure. In the vicinity of a fixed point the RG flow can be expanded in powers of the couplings. 
The information on the universal quantities is encoded in the coefficients of this expansion, in which the linear and the quadratic parts play a special role. 
The linear terms give rise to the so-called critical exponents which are related to the scaling properties of the operators of the theory and have been the focus of most RG studies so far.
The quadratic terms instead give information on some OPE coefficients, which can thus be extracted from the general knowledge of the beta functions.
Whenever scale invariance implies conformal invariance,
the OPE coefficients are directly related to the structure constants of the underlying CFT and thus our analysis
strengthens the link between RG and CFT by showing explicitly how, and which part of, the CFT data
can be determined to some extent directly within an RG approach. 

The scheme dependence of the results can be analyzed in terms of the coupling redefinition connecting two different schemes. It follows that the spectrum is invariant and in this sense universal, while the coefficients of the quadratic terms generally transform inhomogeneously under coupling redefinitions.
In a dimensionally regularized $\overline{\rm MS}$ scheme one has access only to the ``massless'' quadratic coefficients
which are universal at $d=d_c$, but potentially differ in other schemes at higher orders in the $\epsilon$-expansion.
A first observation is that our computational scheme gives the correct values at order $O(\epsilon)$.

After a first pedagogical application of the approach to the investigation 
of the {\tt Ising} universality class, we have introduced a very convenient functional generalization of the standard perturbative RG, in which all the beta functions for the couplings are obtained from few simple generating functions: the beta functionals. 
This functional perturbative framework is a very useful tool that naturally organizes the beta functions in simple generating functionals with few independent c-number coefficients, which, we stress, at leading and next-to-leading order are RG-scheme independent. As a result all the quantities
we have computed (anomalous dimensions at order $O(\epsilon^2)$ and OPE coefficients at order $O(\epsilon)$)  depend essentially only on these universal coefficients.

The simplest of these generating functionals is $\beta_V$, which encodes the RG flow of the whole potential $V(\phi)$
and thus of all couplings of operators of the form $\phi^k$. Contributions of operators involving more derivatives can be included systematically.
The first such contribution comes from $\beta_Z$, which generates all beta functions
of the $O(\partial^2)$ operators $\phi^k (\partial \phi)^2$  included in the field dependent wavefunction $Z(\phi)$.
A goal of our work has been to emphasize some of the advantages of this shift towards a functional approach to standard perturbation theory
in the $\epsilon$-expansion, because it grants an easy and systematic determination of important universal quantities
like both the scaling dimensions and expressions for the massless $\overline{\rm MS}$ OPE coefficients.

As a first application of the \emph{functional} perturbative RG we have reconsidered the {\tt Ising} and the {\tt Lee-Yang} universality classes, as representative of the multicritical unitary and non unitary families,
at the level of the local potential approximation (LPA), i.e.\ without taking into account derivative interactions,
and we have showed how the results extracted from the RG coincide with those recently obtained with CFT techniques. 
This we take as evidence that the $\overline{\rm MS}$ scheme is effective in the computations of the leading
$\epsilon$-corrections to some of the OPE coefficients.

We have also outlined a systematic approach to the inclusion of higher derivative interactions in functional perturbative RG,
and discussed the general mixing patterns among operators at different orders in the derivative expansion
using the general $\phi^{2n}$ models as examples. 
The efficiency of the functional RG techniques is most clearly seen in this context, 
since we are able to collect infinite towers of critical exponents and OPE coefficients at order $O(\epsilon)$ in compact formulas. 
This in particular has allowed a straightforward check of our $\overline{\rm MS}$ estimates
with the results obtained recently with CFT techniques~\cite{Codello:2017qek,Gliozzi:2016ysv}. 
We stress that for the general $\phi^{2n}$ our approach is a multi-loop analysis which, for almost all models, is characterized by an $\epsilon$-expansion
below a \emph{fractional} critical dimension~\cite{Gracey:2017okb}.

The computation of the anomalous dimensions for the multi-critical models in a functional framework was previously carried out 
by O'Dwyer and Osborn~\cite{ODwyer:2007brp}
and here we have limited ourself essentially to the same order of the perturbative expansion.
The analysis is first done without taking into account derivative interactions, and afterwards including the leading order of the mixing
with $O(\partial^2)$ derivative interactions.
Using dimensional analysis we have imposed further constraints on the stability matrix by determining possible terms
that can appear at the quadratic level in any beta function. In particular we have shown that at order $\epsilon$
the stability matrix is lower-triangular. This allows, in agreement with the CFT analysis, the determination
of the anomalous dimensions for all operators contained in the potential $V(\phi)$ and the wavefunction $Z(\phi)$ up to order $\epsilon$, and therefore is not limited to the relevant operators. 
We have also given the $O(\epsilon^2)$ results for the scaling dimensions up to the marginally irrelevant operator $\phi^{2n}$,
and provided the $O(\epsilon)$ values for a family of infinite ``massless'' OPE coefficients
for all the multi-critical universality classes considered, which are found to match the corresponding values from CFT analysis,
when available. 
Clearly more investigations on universality and scheme dependence of the results presented here are required.
Here we just observe that the functional framework in the context of perturbation theory $\epsilon$-expansion strongly constrains the possible redefinitions of the couplings giving support for the success of our approach.
We shall discuss in detail these issues in a forthcoming paper~\cite{CSVZ5}.
On the other hand this fact can be seen as another argument in favor of adopting a functional approach to RG analysis.
To the best of our knowledge, very few results have been obtained even at order $O(\epsilon)$ in CFT computations.
It would be especially interesting to have CFT results at order $O(\epsilon^2)$ whose comparison with
the RG NNLO estimates would show possible artifacts induced by $\overline{\rm MS}$ scheme.

Summarizing, the main results of our paper are highlighted as follows: Inspired by the analysis presented by Cardy~\cite{Cardy:1996xt} which relies on an ultraviolet cutoff, we have proposed to extract the OPE coefficients of a CFT from the coefficients of the quadratic terms in the coupling expansion of the beta functions around the fixed point using dimensional regularization and $\overline{\mathrm{MS}}$ scheme at functional level. We have discussed the scheme dependence of OPE coefficients obtained in this way and identified those that are less sensitive to changes of scheme, which turn out to be the ones that are dimensionless at the upper critical dimension. The order $\epsilon$ OPE coefficients that we have found are compared with the literature on CFT approaches and when available with both methods it is shown that the results always agree. This analysis is done for all multicritical even models, for which the beta functions were obtained in \cite{ODwyer:2007brp}, as well as the Lee-Yang model as a representative of the odd multicritical models for which we have reported the functional betas at NLO \eqref{bvz_LY}. We have demonstrated the power of the functional approach by obtaining compact formulas \eqref{Cijk-LY}, \eqref{ope} encompassing an infinite number of OPE coefficients. Similar formulas are obtained for the order $\epsilon^2$ critical exponents of the relevant operators \eqref{gamma_LY}, \eqref{spectrum}\footnote{The first two instances ($n=2,3$) of \eqref{spectrum} were reported in \cite{ODwyer:2007brp}.}
which, for instance, allow us to verify the shadow relations \eqref{multianom} for all such models in one shot. Finally we have argued that dimensional analysis alone constrains the structure of the beta functions and in particular the stability matrix. 
We have used this information to prove that the formulas for the order $\epsilon$ critical exponents \eqref{gammai} are valid for all the $V$ and $Z$ couplings. 

The functional perturbative RG as introduced in this paper is very general and can be systematically pushed
to higher levels of accuracy by including new families of operators, a fact which is made particularly evident by working at the functional level.
We have recently also successfully applied this method to study~\cite{Codello:2017epp} the non unitary family of multi-critical universality classes described by single scalar field models with odd potentials, whose first elements are the {\tt Lee-Yang} and tricritical {\tt Lee-Yang} ({\tt Blume-Capel}~\cite{vonGehlen:1989yn,Zambelli:2016cbw}), 
for which some CFT results are already available~\cite{Codello:2017qek} finding again full agreement.
We plan to further develop the main ideas and apply the method to other universality classes, e.g. for multifield cases,
as well as to carry on investigations at higher orders in perturbation theory. 

Another extremely important line of investigation, which could possibly overcome the limitations
of the perturbative approach, is to move to one of the non-perturbative functional RG 
frameworks~\cite{Wegner:1972ih, Polchinski:1983gv, Wetterich:1992yh, Morris:1994ie}. 
This step is absolutely non trivial because such approaches are based on massive renormalization schemes,
which often result into a much stronger deformation of the basis of scaling operators (as compared to the {\tt Gaussian} basis),
and make it difficult to establish a direct link to the CFT results. We leave this line of investigation to future research.

\smallskip
\bigskip

\noindent
{\bf Acknowledgments}\\
We would like to thank Hugh Osborn and Slava Rychkov for stimulating correspondence.
O.Z.\ acknowledges support by the DFG under grant No.\ Gi328/7-1.
A.C.\ and O.Z.\ are grateful to INFN Bologna for hospitality and support.

\appendix
\numberwithin{equation}{section}
\section{Perturbative expansion}\label{osborn}

In this appendix we review briefly how perturbative calculations in the functional form are done.
We concentrate on the leading and next to leading order results. We will be brief and closely following \cite{ODwyer:2007brp}.


\begin{figure}[b]
\includegraphics[width=0.35\textwidth]{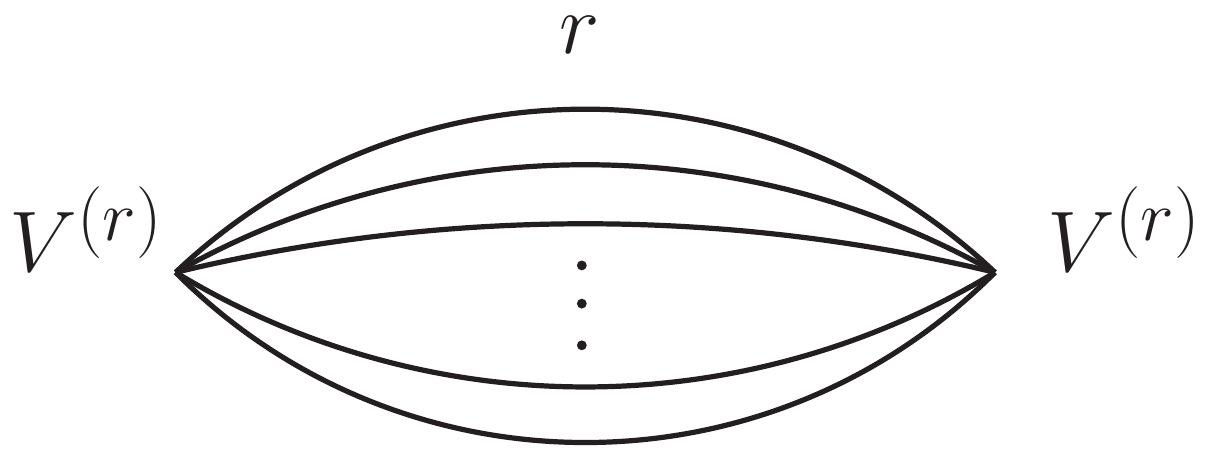}
\caption{Diagram contributing to the counter-term of the potential $V$ and the function $Z$ at quadratic level in the couplings}
\label{melon}
\end{figure}

\begin{figure}
\includegraphics[width=0.42\textwidth]{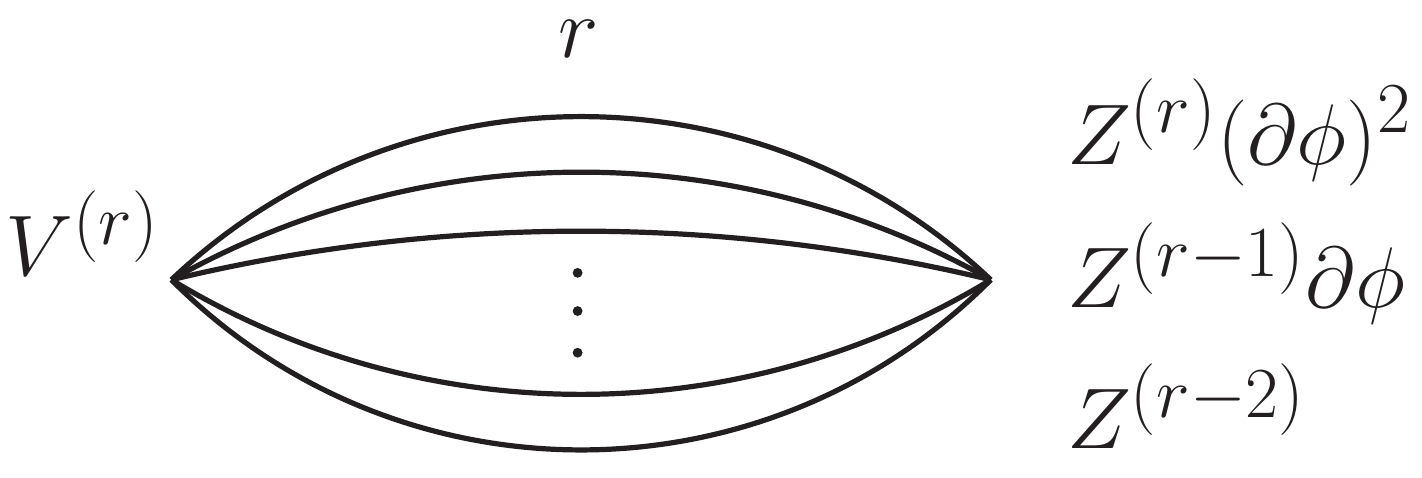}
\caption{Diagrams contributing to the counter-term of the $Z$ function at quadratic level in the couplings}
\label{zmelon}
\end{figure}

The $\epsilon$-expansion is intimately related to an expansion in the couplings through their fixed point value.
The perturbative expansion performed here is therefore in powers of couplings that define the potential $V$ or the function $Z$.
However, in dimensional regularization, given a universality class $\phi^{2n}$  and restricting to operators of a fixed number of derivatives,
there is a one-to-one correspondence between terms of a certain loop order in the beta functional and those of fixed coupling order. 

The leading order counter-terms are quadratic in the couplings. At this level there are two possible terms that contribute to the $V,Z$ counter-terms.
One is represented diagrammatically as in Fig.~\ref{melon} and involves $V$ contributions only. The corresponding expression for this diagram is 
\be  \label{vv}
\sum_{r\geq 2}\,\frac{1}{2\,r!}\int {\rm d}^d x \,{\rm d}^d y  \,V^{(r)}(\phi_x)\,G^r_{x-y}\, V^{(r)}(\phi_y)\,.
\ee  
It turns out that for $r=n$ this ``melon'' type diagram has a pole that contributes to the potential.
On the other hand, for $r=2n-1$ there is a pole term with two derivatives that contributes to the function $Z$.
The corresponding counter-terms in the $\overline{\rm MS}$ scheme can be straightforwardly computed using \eqref{vv} and are given by
\be \label{lct}
S_{c.t.}(\phi) =\frac{1}{\epsilon} \int {\rm d}^d x \left\{\frac{c^{n-1}}{4\, n!}\, V^{(n)}(\phi)^2 -  \frac{(n-1)c^{2n-2}}{16\, (2n)!}\, V^{(2n)}(\phi)^2(\partial\phi)^2\right\}\,. 
\ee
The first counter-term is therefore of $(n-1)$-loop order, while the second term is at $2(n-1)$ loops.
The other diagram that contributes at quadratic level is shown in Fig.~\ref{zmelon}.
This involves both the $V$ and $Z$ functions and contributes to the flow of $Z$ for $r=n$,
which will therefore be of $(n-1)$-loop order. Notice that there are three different diagrams of this kind depending on whether one,
two or none of the fields in $(\partial\phi)^2$ are involved in the propagators, as shown in Fig.~\ref{zmelon}.

\begin{figure}[h]
\begin{subfigure}[a]{0.4\textwidth}
\includegraphics[width=\textwidth]{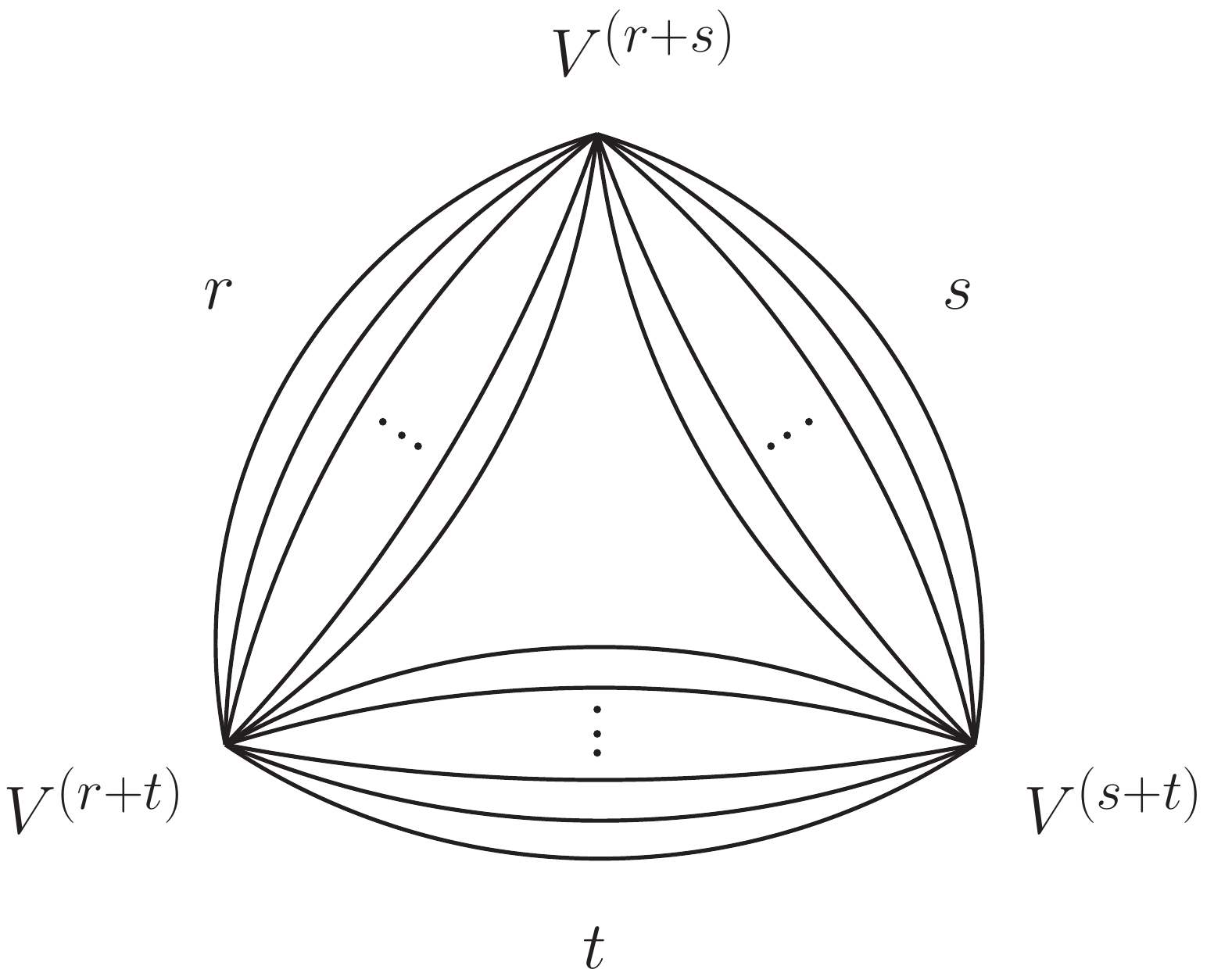}
\caption{} \label{a}
\end{subfigure}
\raisebox{-80pt}{
\begin{subfigure}[b]{0.55\textwidth}
\includegraphics[width=\textwidth]{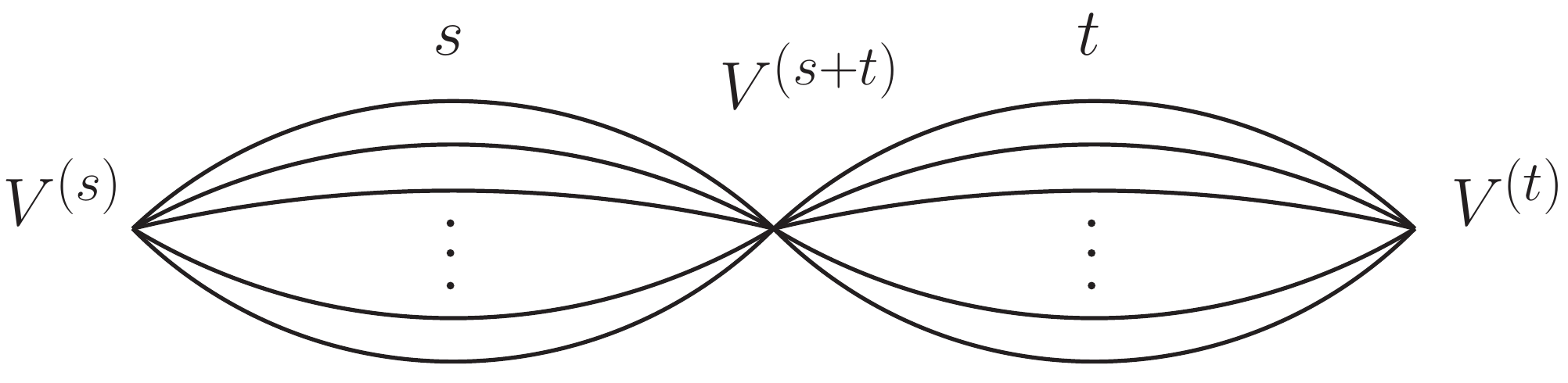} \vspace{-28pt}
\caption{} \label{b}\vspace{8pt}
\includegraphics[width=0.63\textwidth]{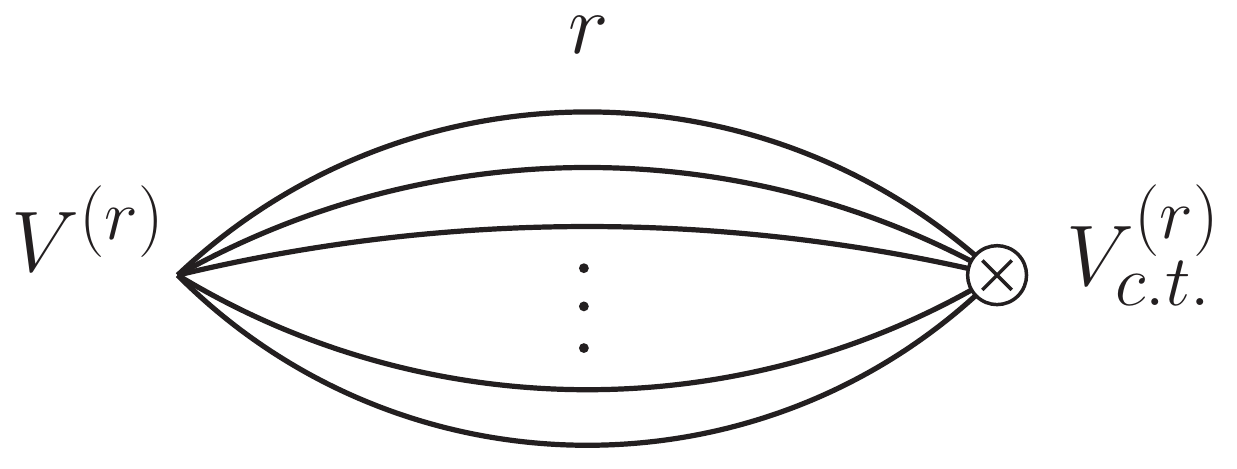} \vspace{-3pt}
\caption{} \label{c}
\end{subfigure}}
\caption{Diagrams contributing to the counter-term of the potential at cubic level in the couplings.}
\end{figure}

At cubic order in the couplings, restricting to the contribution from $V$ only, i.e. LPA,
there are three types of counter-term diagrams for the potential. The first one can be seen as
a one loop graph with three vertices whose propagators are replaced with a bunch of $r$, $s$ and $t$ propagators as shown in Fig.~\ref{a}.
In order to have a pole contributing to the potential the number of propagators must be constrained to $r+s+t=2n$.
The second one consists of two melon diagrams as in Fig.~\ref{b}, and the third graph, shown in Fig.~\ref{c},
is a melon diagram involving the potential and its counter-term at quadratic level $V_{c.t.}(\phi)$, which is the first term on the right-hand side of \eqref{lct}. 
In both diagrams the $\epsilon$ singularity that contributes to the potential occurs when the number of propagators in each melon is equal to $n$. 
These three diagrams are therefore all of $2(n-1)$-loop order. They give rise to the cubic terms in the second and third lines of \eqref{bv}.

The precise relation between the counter-terms and the dimensionful beta functions of the potential at quadratic level $\beta_{V,2}$ and at cubic level $\beta_{V,3}$ and also the dimensionful beta of the wavefunction at quadratic level $\beta_Z$ are given by the following equations
\be 
\beta_{V,2} = \epsilon V_{c.t.2} - \left.\mu\frac{d}{d\mu}\right|_1\!\!V_{c.t.2}= (n-1)\epsilon V_{c.t.2}, 
\ee
\be 
\beta_{V,3} = \epsilon V_{c.t.3} - \left.\mu\frac{d}{d\mu}\right|_1\!\!V_{c.t.3} - \left.\mu\frac{d}{d\mu}\right|_2\!\!V_{c.t.2} = 2(n-1)\epsilon V_{c.t.3} - \left.\mu\frac{d}{d\mu}\right|_2\!\!V_{c.t.2},
\ee
\be 
\beta_Z = - \left.\mu\frac{d}{d\mu}\right|_1\!\!Z^{v^2}_{c.t.2}- \left.\mu\frac{d}{d\mu}\right|_1\!\!Z^{vz}_{c.t.2}= 2(n-1)\epsilon Z^{v^2}_{c.t.2}+(n-1)\epsilon Z^{vz}_{c.t.2}.
\ee
The total beta of the potential $\beta_V = \beta_{V,2}+\beta_{V,3}$ and the beta of the wavefunction $\beta_Z$ are related to the dimensionless betas through \eqref{betav} and \eqref{betaz}, respectively. The cubic counter-term $V_{c.t.3}(\phi)$ is the sum of the diagrams \eqref{a}, \eqref{b} and \eqref{c}, and the quadratic counter-terms for the wavefunction $Z^{v^2}_{c.t.2}$ and $Z^{vz}_{c.t.2}$ are extracted respectively from the second term on the r.h.s of \eqref{lct} and from the counter-term diagram of Fig.~\eqref{zmelon} for $r=n$. The $\mu$-derivatives with an index $1$ are taken using the tree-level flow, which for the derivatives of the potential and the wavefunction are given by the following relations
\be \label{lf}
\left.\mu\frac{d}{d\mu}\right|_1\!\!V^{(r)} = -\frac{r-2}{2}\,\epsilon\, V^{(r)}, \qquad 
\left.\mu\frac{d}{d\mu}\right|_1\!\!Z^{(r)} = -\frac{r}{2}\,\epsilon\, Z^{(r)},
\ee
while the $\mu$-derivative with an index $2$ is based on the quadratic flow. In particular
\be \label{qf}
\left.\mu\frac{d}{d\mu}\right|_2\!\!V^{(r)} =  \beta_{V,2}^{(r)}.
\ee

\section{A general scaling relation}\label{genscaling}

In this appendix we would like to obtain a relation valid among the scaling of two couplings induced by the RG flow.
This information can then be compared to the relation obtained in CFT for the scaling of the field operator and one of its descendants.

We have already encountered the scaling dimensions of the operator $\phi^i$ and its corresponding dimensionless coupling $g_i$, 
which were denoted by $\Delta_i$ and $\theta_i$ in Eqs.~\eqref{Delta} and~\eqref{theta}, 
together with their anomalous parts $\gamma_i$ and  $\tilde\gamma_i$ respectively.
Let us consider for a moment the case of a multi-critical theory $\phi^{2n}$. Then, for $i\neq 2n-1$, the relation $\theta_i+\Delta_i = d$ holds. 
This is equivalent to $\gamma_i=\tilde\gamma_i$. Instead, for the descendant operator corresponding to $i=2n-1$ this relation is modified to $\theta_i+\Delta_i = d+\eta$ 
by the presence of $\eta=2\gamma_1=2\tilde\gamma_1$, which is twice the anomalous dimension of $\phi$. 
One can link this fact to the relation $\gamma_{2n-1}=(n-1)\epsilon +\gamma_1$ coming from the descendant constraint in CFT, $\Delta_{2n-1}=2+\Delta_1$ 
and from another relation that we shall prove in general in the following. 
Indeed we shall see that the latter is equivalent to $\tilde\gamma_{2n-1}+\tilde{\gamma}_1=(n-1)\epsilon$ so that the two anomalous dimensions
(associated to the CFT operator and RG coupling) are related by $\gamma_{2n-1}=\tilde\gamma_{2n-1}+\eta$.

We shall work at a general functional level~\cite{Wegner:1972ih}.
Let us consider for the truncation with two functions $V$ and $Z$ 
which describes deformations with composite (non-total derivative) operators containing up to two derivatives. 
The beta functions describing the RG flow  are generically written as in Eqs.~\eqref{betav} and~\eqref{betaz}.

Linearizing such equations around the FP one obtains
\bea
 \theta \delta v &=&\sum_i \frac{\partial \beta_v}{\partial v^{(i)}} \delta v^{(i)} \!
 +\! \sum_i \frac{\partial \beta_z}{\partial z^{(i)}} \delta z^{(i)}
 \\&=& -d\, \delta v + \frac{1}{2}(d\!-\!2\!+\!\eta)\varphi \,\delta v' 
 +\mu^{-d} \sum_i \frac{\partial \beta_V}{\partial v^{(i)}} \delta v^{(i)}+
 \mu^{-d} \sum_i \frac{\partial \beta_V}{\partial z^{(i)}} \delta z^{(i)} \nn \\
 \theta \delta z &=& \sum_i \frac{\partial \beta_z}{\partial v^{(i)}} \delta v^{(i)} \!+\! \sum_i \frac{\partial \beta_z}{\partial z^{(i)}} \delta z^{(i)}\\&=&
 \eta\, \delta z + \frac{1}{2}(d\!-\!2\!+\!\eta)\varphi\, \delta z'+ Z_0^{-1}\sum_i \frac{\partial \beta_Z}{\partial v^{(i)}} \delta v^{(i)}+ 
 Z_0^{-1} \sum_i \frac{\partial \beta_Z}{\partial z^{(i)}} \delta z^{(i)}\nn\,.
\eea
Let the fixed point solution be $(v_*(\varphi),z_*(\varphi))$.
Taking the derivative in $\varphi$ of the fixed point equations
\bea
0 &=& \frac{d \beta_v(v_*, z_*,\varphi)}{d \varphi}= \sum_i \frac{\partial \beta_v}{\partial v^{(i)}}v_*^{(i+1)}+ \sum_i \frac{\partial \beta_z}{\partial z^{(i)}}  z_*^{(i+1)}
+\frac{\partial \beta_v}{\partial \varphi}
\nn \\
0 &=& \frac{d \beta_z(v_*, z_*,\varphi)}{d \varphi}= \sum_i \frac{\partial \beta_z}{\partial v^{(i)}} v_*^{(i+1)}+ \sum_i \frac{\partial \beta_z}{\partial z^{(i)}} z_*^{(i+1)}
+\frac{\partial \beta_z}{\partial \varphi}\,,
\eea
one immediately sees that $(\delta v,\delta z)_{r}=(v'_*(\varphi),z'_*(\varphi))$ is a solution of the linearized equation and is a relevant eigenoperator with eigenvalue $\theta_r=\frac{1}{2}(d-2+\eta)$.
Moreover, since $\sum_i \frac{\partial \beta_{V,Z}}{\partial v^{(i)}} \delta v^{(i)}$ contains only terms with at least two derivatives on $v$,
one can easily check that $(\delta v,\delta z)_1=(\varphi, 0)$ is a solution of the linearized equations and is a relevant eigenoperator with eigenvalue $\theta_1=\frac{1}{2}(d+2-\eta)$.
Therefore one can immediately obtain from the RG flow the scaling relation $$\theta_r+\theta_1=d\,.$$
Specializing now to the multi-critical $\phi^{2n}$ models, this is equivalent to the relation given after Eq.~\eqref{multianom}, i.e.\ $\tilde{\gamma}_{2n-1}+\tilde{\gamma}_1=(n-1)\epsilon$.


\section{Relations with the functional non-perturbative RG}\label{relFRG}

In this appendix we want to spell out an interesting relation that the functional perturbative RG has with the functional non-perturbative RG
in the effective average action implementation that was originally proposed by Wetterich~\cite{Wetterich:1992yh} and independently by Morris~\cite{Morris:1994ie}.

In this approach a scale $k$ is introduced by modifying the theory's propagator through the inclusion of an IR cutoff $R_k$ in momentum space.
This modification generates an RG flow equation for the generator of the irreducible diagrams
\begin{equation}
 \begin{split}
  k\partial_k \Gamma_k &= \frac{1}{2} \Tr \left(\Gamma^{(2)}+R_k\right) k\partial_k R_k \,.
 \end{split}
\end{equation}
Using a truncation of the space of all possible operators appearing in $\Gamma_k$ such as \eqref{LPA} and adopting a specific form for the cutoff,
we can compute the flow of the effective potential
\begin{equation}\label{betaV-nprg}
  \beta_V = k\partial_k V  = c_d \frac{k^{d+2}}{k^2+V''} \,,
\end{equation}
in which we defined $ c_d^{-1} = (4\pi)^{d/2}\Gamma(1+d/2)$.
To compute the above flow one can choose to work with the so-called optimized cutoff $R_k(q^2) = (k^2-q^2)\theta(k^2-q^2)$
because the result is particularly simple, but the results of this appendix will be independent of this particular choice.

Let us expand the right hand side of \eqref{betaV-nprg} in powers of $V''(\varphi)$
\begin{equation}
\begin{split}\label{betaV-nprg-expanded}
  \beta_V & = c_d \Bigl\{ k^d - k^{d-2} V'' + k^{d-4} (V'')^2 -k^{d-6} (V'')^3 +\dots \Bigr\}\,.
\end{split}
\end{equation}
For any given dimensionality, we shall refer to the terms of this expansion as \emph{critical} if they scale as $k^0$ and \emph{off-critical} if they do not.
For example the term $c_d k^{d-4}(V'')^2 $ is critical in $d=4$, while all other terms are off-critical.
The critical terms have two important properties:
On the one hand they are independent by the cutoff; this is because once the momentum scale $q^2$ is integrated out, the scale $k$ is what remains of $R_k(q^2)$,
so independence of $k$ implies independence of the cutoff function itself (this of course can be proven more rigorously).
On the other hand they are related to the logarithmic divergences of the theory; using again $d=4$ as an example
\begin{equation}
  V = - \int^\Lambda \frac{{\rm d}k}{k} \beta_V \sim - c_d \int^\Lambda \frac{{\rm d}k}{k} k^{d-4} (V'')^2  \sim -(V'')^2 \log \Lambda  \qquad {\rm in}\quad d=4\,,
\end{equation}
which also implies that they correspond to the $\frac{1}{\epsilon}$ poles of dimensionally regulated perturbation theory.

It is instructive to choose a procedure that deliberately removes the off-critical terms from the flow \eqref{betaV-nprg}.
We obtain
\begin{eqnarray}
  \beta_V &=&  \phantom{-}c_4  (V'')^2  =  \phantom{-}\frac{1}{2(4\pi)^2}(V'')^2   \quad {\rm in }\quad d=4 \\
  \beta_V &=& -c_6  (V'')^3  = -\frac{1}{6(4\pi)^3}(V'')^3   \quad {\rm in }\quad d=6\,.
\end{eqnarray}
It is easy to see that the above results correspond to the leading one loop contributions of the two tutorial examples {\tt Ising} and {\tt Lee-Yang}.
These two examples are the only two universality classes that are captured through critical terms by the above procedure,
even though \eqref{betaV-nprg} is well known to be able to ``see'' critical points corresponding to all the $\phi^{2n}$ models \cite{Codello:2012sc} and more \cite{Zambelli:2016cbw}. 
The reason why only those two critical terms appear has to do with the fact that a local potential truncation of the operator space of $\Gamma_k$
does not contain all possible terms that can be generated perturbatively by higher loops.
This should also explain why \eqref{betaV-nprg} returns only the leading terms of the {\tt Ising} and {\tt Lee-Yang} universality classes.
The study of truncations that include the higher loops effects has been initiated in \cite{Codello:2013bra}, in which also the scheme-dependence of functional renormalization group is carefully investigated,
but those results have not yet been formulated in a fully functional form
as in the models of the present paper.

The careful reader must have noticed that the second term of \eqref{betaV-nprg-expanded} is critical for any even value of $d$.
In $d=2$ the critical model corresponds to the {\tt Sine-Gordon} universality class. The beta function of the dimensionful potential is
\begin{eqnarray}
  \beta_V &=&  -c_2  V''  =  -\frac{1}{4\pi}V'' \,,  \qquad {\rm in }\quad d=2\,.
\end{eqnarray}
It is interesting to investigate explicitly the flow of the dimensionless potential in $d=2$, which is
\begin{eqnarray}\label{sine-gordon-dimless}
  \beta_v &=& -2v(\varphi) -\frac{1}{4\pi}  v''(\varphi) \,.
\end{eqnarray}
The above beta function does not contain the scaling term contributed by the field $\varphi$ because the field is canonically dimensionless in $d=2$
and fluctuations do not generate a nonzero anomalous dimension.
Interestingly, the fixed point solution of the {\tt Sine-Gordon} universality can be obtained by directly integrating the right hand side of \eqref{sine-gordon-dimless}.
Using $v''(0)=\sigma$ as boundary condition we obtain
\begin{eqnarray}
  v(\varphi) &=&  -\frac{\sigma}{8\pi}  \cos (\sqrt{8\pi} \varphi) \,,
\end{eqnarray}
in which we can recognize the well-known Coleman phase $\sqrt{8\pi}$.
This fact is quite amazing since the Coleman phase is a non-perturbative result, which we just obtained on the basis of a perturbative approximation.
We plan to return to the study of the {\tt Sine-Gordon} universality class and of all other universal terms in a future work.


As mentioned above, the method presented in this appendix is limited to the universal terms which come from one-loop diagrams because of the local potential truncation.
The truncation of this appendix is by definition unable of dealing with higher derivative operators, or operators which are generally generated
beyond the first loop. Furthermore, we have made a specific choice of the cutoff which forces us to resort to the rather brute force method of ``chopping'' all nonzero powers of the cutoff scale $k$
to locate universal terms. A more refined approach to both these shortcomings which also aligns with our discussion of the scheme transformations of Sect.~\ref{transformation}
can be found in \cite{Lizana:2017sjz} where special ``normal'' coordinates in the space of all couplings are found in the context of the functional renormalization group
(using the Polchinski equation instead of the Wetterich equation, but arguably the conclusions are very similar).
The normal coordinates of \cite{Lizana:2017sjz} could be understood as a geometrical generalization of the basis of couplings with well-behaved scaling properties introduced in Sect.~\ref{transformation}
and their application clearly shows that a consistent renormalization of correlators of all composite operators, thus including in principle all possible OPE coefficients,
is possible within the functional renormalization group approach (at least in the vicinity of the Gaussian fixed point).
In order to achieve the same results, the functional method presented in the main text of this paper requires the consistent inclusion of higher derivative operators
according to their mixing patters as described in Sect.~\ref{sec de}.


\end{document}